\documentclass[11pt]{amsart}
\usepackage[utf8]{inputenc}
\usepackage{amsmath,amssymb,amsthm,mathtools}
\usepackage{enumitem}
\usepackage{algorithm}
\usepackage[noend]{algpseudocode}
\usepackage{booktabs}
\usepackage[margin=1in]{geometry}
\usepackage{xcolor}
\usepackage[hidelinks]{hyperref}
\usepackage{cite}
\usepackage[normalem]{ulem}

\usepackage{tikz}
\usepackage{orcidlink}

\newtheorem{theorem}{Theorem}[section]
\newtheorem{lemma}[theorem]{Lemma}
\newtheorem{proposition}[theorem]{Proposition}
\newtheorem{corollary}[theorem]{Corollary}
\theoremstyle{definition}
\newtheorem{definition}[theorem]{Definition}
\newtheorem{example}[theorem]{Example}
\newtheorem{remark}[theorem]{Remark}

\newcommand{\F}{\mathbb{F}}
\newcommand{\Z}{\mathbb{Z}}
\newcommand{\N}{\mathbb{N}}
\newcommand{\wt}{\mathrm{wt}}
\newcommand{\wtH}{\mathrm{wt}_H}
\newcommand{\wtl}{\mathrm{wt}_\lambda}
\newcommand{\dH}{d_H}
\newcommand{\dl}{d_\lambda}
\newcommand{\ev}{\mathrm{ev}}
\newcommand{\Tr}{\mathrm{Tr}}
\newcommand{\Gap}{\mathrm{Gap}}
\newcommand{\Amp}{\mathrm{Amp}}
\newcommand{\cD}{\mathcal{D}}
\newcommand{\bcD}{\overline{\mathcal{D}}}
\newcommand{\card}[1]{\#{#1}}
\newcommand{\Clos}[2]{{#1}_{#2}^{\supseteq}}
\newcommand{\Int}[2]{{#1}_{#2}^{\subseteq}}
\newcommand{\cA}{\mathcal{A}}
\newcommand{\cI}{\mathcal{I}}
\newcommand{\cJ}{\mathcal{J}}
\newcommand{\subfield}[2]{{#1}_{#2}}
\newcommand{\sfCq}[2]{\mathcal{C}_{#1,#2}}
\newcommand{\sfC}[1]{\mathcal{C}_{#1}}

\title[Asymmetric quantum codes in the subfield metric]{Asymmetric quantum codes in the subfield metric}
\author[C. Galindo]{Carlos Galindo \orcidlink{0000-0002-3908-4462}}
\author[F. Hernando]{Fernando Hernando \orcidlink{0000-0002-9758-2152}}
\address[Carlos Galindo and Fernando Hernando]{Instituto
Universitario de Matem\'aticas y Aplicaciones de Castell\'on and
Departamento de Matem\'aticas, Universitat Jaume I, Campus de Riu
Sec, 12071 Castell\'o, Spain}
\email{galindo@uji.es}
\email{carrillf@uji.es}

\author[D. Ruano]{Diego Ruano \orcidlink{0000-0001-7304-0087}}
\address[Diego Ruano]{IMUVA-Mathematics Research Institute and,
Universidad de Valladolid, 47011 Valladolid, Spain}
\email{diego.ruano@uva.es}

\author[R. San-Jos\'e]{Rodrigo San-Jos\'e \orcidlink{0000-0002-0944-7584}}
\address[Rodrigo San-Jos\'e]{Department of Mathematics,
Virginia Tech, Blacksburg, VA, USA.}
\email{rsanjose@vt.edu}
\date{}

\begin{document}
\begin{abstract}
The $\lambda$-subfield metric applied to a classical code $C \subseteq \mathbb{F}_{q^m}^n$ was introduced by Grassl, Horlemann, and Weger. It distinguishes the coordinates of a vector over $\mathbb{F}_{q^m}$ that lie in the base field $\mathbb{F}_q$ from those that do not, making it a natural tool for asymmetric quantum error correction, where phase-shift errors correspond to the base field. We prove a general lower bound for the minimum $\lambda$-subfield distance of a linear code $C$ in terms of the minimum distances of $C$, its subfield subcode $\subfield{C}{q}$, and the $\mathbb{F}_q$-linear code $\{c-c^q : c \in C\}$. For cyclic codes  these three distances are bounded by gaps of explicit defining sets. We use this fact to construct Hermitian dual-containing cyclic codes over $\mathbb{F}_{q^2}$ of length $q^{2s}-1$ (and, more generally, of every length $n$ such that the order of $q$ modulo $n$ is even) whose subfield subcodes have designed distances up to an explicit bound, which we show cannot be exceeded. Furthermore, we determine the largest possible dimension of such codes and provide a procedure to improve the term of the bound arising from $\{c-c^q:c\in C\}$. 
The obtained lower bound for the $\lambda$-subfield distance shows that the resulting quantum codes correct many more phase-shift errors than arbitrary errors and that, for suitable values of $\lambda$, some of them have a larger $\lambda$-subfield distance than every pure asymmetric CSS code of the same length and dimension.
\end{abstract}
\maketitle

\section{Introduction}

Recent experimental demonstrations of quantum advantage using programmable superconducting processors~\cite{Aru} and photonic boson-sampling devices~\cite{Zhong} have underscored the transformative potential of quantum information
processing across science and industry.
Realizing this potential at scale requires shielding quantum states from
the decoherence induced by unavoidable interactions with the environment.
The no-cloning theorem~\cite{Woot} precludes the na\"{\i}ve redundancy
strategies of classical computing, making dedicated quantum
error-correcting codes (QECCs) an indispensable ingredient of any
scalable quantum architecture.

The theory of QECCs was initiated by Shor~\cite{shorFirstQECC} and
Steane~\cite{css2,cssoriginal2} and placed on a firm algebraic footing by
Calderbank, Rains, Shor and Sloane~\cite{css1,Calder2}; early
investigations concentrated on binary systems~\cite{calderbankgoodquantum,Gottesman}.
The central result---the stabilizer formalism~\cite{Got97}---allows one
to construct QECCs from classical codes that are self-orthogonal with
respect to a suitable inner product~\cite{css1,calderbankgoodquantum,css2,kkks}.
Research subsequently extended the framework to $q$-ary codes for any
prime power~$q$, exploiting the richness of classical coding
theory---BCH codes, Reed--Solomon codes, algebraic-geometric codes---for
quantum purposes~\cite{AshKni01,GBR04,galindodistancestabilizer,FLX}, and producing a wide
variety of quantum codes through diverse
methodologies~\cite{bierbrauerquantum,AKS,kkks,XingC,LaG12,Lag2,gahe,relativehull,sanjose_good_triorthogonal}.

In many physical implementations, however, the two fundamental types of errors involved in quantum correction are far from
equally likely: phase-shift ($Z$-type) errors are typically much more frequent than qudit-flip
($X$-type) errors. This asymmetry, well documented in nuclear magnetic resonance,
superconducting and quantum-dot qubits~\cite{ioffe}, motivates the study of \emph{asymmetric} quantum codes,
which sacrifice correction power against the less likely error type in order to protect much
more strongly against the more likely one. The CSS (Calderbank--Shor--Steane) construction
\cite{calderbankgoodquantum,css2} naturally produces asymmetric codes: denoting by $\mathrm{wt}_H$ (respectively, $d_H$) the Hamming weight (respectively, distance) and by $\perp$ the Euclidean dual, two nested classical codes
$C_2\subseteq C_1$ over $\F_q$ of length $n$ yield an
$[[n,\dim C_1-\dim C_2,d_Z/d_X]]_q$ code, where $d_Z=\mathrm{wt}_H\left(C_2^\perp \setminus C_1^\perp\right)$ governs phase-shift correction and
$d_X=\mathrm{wt}_H(C_1 \setminus C_2)$ governs qudit-flip correction. When $d_Z = d_H\left(C_2^\perp\right)$ and $d_X =d_H(C_1)$, the resulting code is said to be pure. This approach has been intensively studied
\cite{asymmetricConstructions,WFLX,EJLP,LaG12,Liu,christensennested,Camps-Moreno2025-xo,galindoimprovednestedcodepairs}, and many of the best known asymmetric codes come from it, although additive (non-CSS) asymmetric codes have also been studied~\cite{ELS,EzermanGrassl}.

In some recent architectures, the asymmetry between phase flips and bit flips is engineered
rather than merely observed. In bosonic cat qubits, bit flips are suppressed exponentially in
the mean photon number, while phase flips grow only linearly~\cite{GM,Puri}. This is the
principle behind the Ocelot chip of Amazon Web Services~\cite{Putterman}, which concatenates
five cat qubits with a repetition code that corrects phase flips, and behind the LDPC-cat codes
developed by researchers at Alice~\&~Bob and Inria~\cite{LDPCcat}, which replace the repetition
code by a classical LDPC code to reduce the overhead. In both cases the residual bit flips are
left uncorrected, so their contribution grows with the length of the code, and suppressing them
further costs additional phase flips. The resulting noise is strongly biased, and quantum codes that protect mainly against phase flips while still handling
occasional bit flips are a natural complement.

Grassl, Horlemann and Weger~\cite{GHW} proposed a different
viewpoint. Identifying the error operators on a qudit with the elements of $\F_{q^2}$, the
diagonal (phase-shift) errors correspond to the subfield $\F_q$, and the remaining errors to
$\F_{q^2}\setminus\F_q$.

In~\cite{GHW}, the more general setting of extension fields $\F_{q^m}$, $m>1$, is considered. This leads to the \emph{$\lambda$-subfield weight} on $\F_{q^m}$,
which assigns weight $1$ to the nonzero elements of $\F_q$ and weight $\lambda \geq \frac{1}{2}$ to the
elements of $\F_{q^m}\setminus\F_q$. A code with large $\lambda$-subfield distance corrects
many phase-shift errors together with fewer general errors. 
The distinction (diagonal versus
non-diagonal errors) is different from the $X/Z$ distinction of the CSS approach: it works
directly at the level of field elements, treating $X$ and $Y$ errors alike, instead of handling
the $X$-part and the $Z$-part of an error separately (the CSS approach treats a $Y$-error as the combination of an $X$- and a $Z$-error).
The basic theory of the metric is developed in~\cite{GHW}: bounds, the asymptotic
behavior of random codes and MacWilliams identities for $m=2$ (a linear programming bound for a family of partitions containing the subfield metric is studied in~\cite{BCW}). In~\cite{GHW} it is also shown that a family of classical codes (coming from MRD codes) attains a Singleton-type bound for this metric. More recently, Kosovac, Plechinger, Spirovski, and Weger~\cite{KPSW} studied maximum $\lambda$-subfield distance codes, their asymptotic densities, and constant-weight codes, and established Griesmer-, Elias-, and Johnson-type bounds. However, explicit families of \emph{Hermitian dual-containing} codes with good $\lambda$-subfield distance, which are what we consider to construct good quantum codes, are missing, and so is a precise comparison with asymmetric CSS codes: Grassl, Horlemann and Weger give two examples of cyclic codes of length $17$~\cite[Examples~12 and~13]{GHW} and leave the construction of applicable quantum codes for future work.

In this paper we address this problem by considering \emph{cyclic codes}, using their description as evaluation codes of
polynomials in one variable $X$ over $\F_{q^r}$, $r$ being a multiple of $m$, evaluated at the $n$-th roots of unity, $n\mid q^r -1$.
BCH codes appear when the complement of the defining set contains a consecutive interval (see Remark~\ref{rem:defset}). Our focus differs from~\cite{GHW,KPSW} because we derive lower bounds for
$\dl(C)$ in terms of classical Hamming distances, and we construct explicit infinite families
of codes with large $\lambda$-subfield distance by exploiting the rich structure of cyclic codes.

We also make precise in which sense the $\lambda$-subfield metric is the right tool. If phase-shift errors occur
with probability $p_Z \neq 0$ and errors involving a qudit-flip with probability $p_X \neq 0$, all non-diagonal errors being equally likely (under the assumptions of Proposition~\ref{prop:channel}), then the probability of an error
pattern is a decreasing function of its $\lambda$-subfield weight for
$$\lambda=1+\ln(q\,p_Z/p_X)/\ln\big((q-1)(1-p_Z-p_X)/p_Z\big),$$ so that the most likely errors compatible with a syndrome are those of minimum $\lambda$-weight (Section~\ref{sec:channel}). For realistic values of $p_Z$ and $p_X$, this gives $\lambda$ between
$1.3$ and $4.2$. The $\lambda$-subfield distance is a finer figure of merit that
applies to every stabilizer code: it always lies between $\min\{d_Z,\lambda d'_X\}$ and $d_Z$, where $d'_X$ is the least number of non-diagonal entries of a nontrivial logical operator, and CSS codes satisfy $d_\lambda=\min\{d_Z,\lambda d_X\}$, while non-CSS codes can do better because every logical operator with few non-diagonal entries may be
forced to carry many diagonal ones, a coupling that the pair $(d_Z,d_X)$ does not see. This is what the codes of this
paper exploit. Non-CSS codes tailored to biased noise, such as the surface codes of~\cite{TBF,XZZX}, also exploit this coupling in the topological setting; here we study algebraic families. Note that for pure dephasing ($p_X=0$), no quantum code of
this type can beat the best classical code, and for weak bias, $\lambda$ is close to $1$ and the best symmetric codes are
preferable. The advantage appears in the intermediate, physically common, regime where phase-shift errors dominate but
qudit-flips are not negligible.

Next, we list our main contributions. To start, we show the channel interpretation of the metric, establish the inequalities $$\min\{d_Z,\lambda d'_X\}\le d_\lambda \le d_Z$$ for stabilizer codes (and the equality $d_\lambda=\min\{d_Z,\lambda d_X\}$ for CSS codes; Proposition~\ref{prop:dzdx}), and state a comparison theorem: if $C$ is Hermitian dual-containing, $k_Q=2\dim_{\F_{q^2}}C-n>0$ (the dimension of the quantum code), $d=\dH(\subfield{C}{q})$, $\dH(C)\ge2$ and there is no $[n,k_Q+1,\ge d]_q$ linear code, then for every $1\le\lambda<d$ the resulting quantum code has a larger $\lambda$-subfield distance than every CSS code of the same length and dimension which is pure in $Z$ and has phase distance at least $d$ (Theorem~\ref{thm:dominance}).

One of our main results is Theorem~\ref{thm:main} which provides a general lower bound for a linear code $C\subseteq\F_{q^m}^n$ and $\lambda \ge 1$:
\begin{equation}
\label{cota}
  \dl(C)\ge\min\{(\lambda-1)\dH(\cD)+\dH(C),\ \dH(\subfield{C}{q})\},
\end{equation}
where $\subfield{C}{q}=C\cap\F_q^n$ is the subfield subcode and $\cD=\{c-c^q:c\in C\}$. In
particular, if $\subfield{C}{q}\neq\{0\}$, then $\dl(C)=\dH(\subfield{C}{q})$ for $\lambda$ large enough
and, for Hermitian dual-containing codes, the dimension of the
quantum code is at most $\dim_{\F_q}\subfield{C}{q}$ (Remark~\ref{rem:quantumbound}). Denoting by $\Tr$ the trace map from $\mathbb{F}_{q^m}$
to $\mathbb{F}_{q}$, we also show that $\dH(\cD)\ge\dH(\Tr(C))$, with equality when $m=2$ (Proposition~\ref{prop:trace}), that for
Hermitian dual-containing codes $(\subfield{C}{q})^\perp\subseteq\Tr(C)$, which limits $\dH(\cD)$
(Proposition~\ref{prop:tradeoff}), and we give an exact expression for $\dl(C)$ and a way of computing it as the
Hamming distance of an $\F_q$-linear code (Propositions~\ref{prop:profile} and~\ref{prop:image}).

We study the subfield metric in the case of univariate evaluation codes $\mathcal{C}_{\mathcal{I}}$ ($\{1\}$-affine variety codes \cite{galindostabilizer}) and cyclic codes $\mathcal{C}_{\mathcal{I},q^m}$ (Definition~\ref{1afine}), which are their subfield subcodes. The main result in Section~\ref{sec:cyclic} is Theorem~\ref{thm:cyclic}, which gives a lower bound for the $\lambda$-subfield distance of a cyclic code $\sfCq{\cI}{q^m}$ in terms of the gaps of the sets $\cI$, $\cI\cup q\cI$ and $\Int{\cI}{q}$.

In Section~\ref{sec:hermitian}, Theorem~\ref{thm:construction} proves that, for $n=q^{2s}-1$, Algorithm~\ref{alg:delta} allows us to construct Hermitian dual-containing cyclic codes
$\sfCq{\cI}{q^2}$$\subseteq\F_{q^2}^n$ such that the subfield subcode $\sfCq{\cI}{q}$ over the field $\mathbb{F}_q$ has designed distance $\delta$, for
every $\delta\le\gamma=(q^{2s}-1)/(q^\tau+1)$, with $\tau$ the largest odd divisor of $s$. The bound $\gamma$ cannot be exceeded by the designed (BCH) distance of the subfield subcode of any Hermitian dual-containing cyclic code of length $q^{2s}-1$ (Corollary~\ref{cor:limit}). 
Moreover, a variant of the construction attains the largest possible dimension, so that the quantum dimension equals the dimension of the $q$-ary BCH code of designed distance $\delta$ (Proposition~\ref{prop:optimal}). 
Theorem~\ref{thm:generaln} extends the construction to every length $n$ such that the order of $q$ modulo $n$ is even. We also give a procedure to enlarge $\dH(\cD)$ in the bound (\ref{cota}), see Algorithm~\ref{alg:enlarge} and Proposition~\ref{prop:enlarge}.

Finally, Section~\ref{sec:hermitian} also contains examples and tables, with the true distances computed with Magma
\cite{magma} and exact values of $\dl(C)$ for the classical code $C$. For instance, for $\lambda=3$ our $[[63,45]]_2$ code corrects every pattern of up to three phase-shift errors and, separately, every single error involving a qubit-flip, which no pure CSS $[[63,45]]_2$ code can do (Example~\ref{ex:63}).

The paper is organized as follows. Section~\ref{sec:prelim} contains notation and
preliminaries on quantum codes, asymmetric codes and the $\lambda$-subfield metric.
In Section~\ref{sec:bound} we prove the general lower bound and its refinements, and a reduction of the decoding problem to decoding in the Hamming metric (Proposition~\ref{prop:decoding}). Section~\ref{sec:channel}
discusses the biased channel and compares the subfield approach with asymmetric CSS codes.
Section~\ref{sec:cyclic} is devoted to cyclic codes: their description and the bound
for these codes. Section~\ref{sec:hermitian} contains the Hermitian dual-containing construction, its optimality,
its extension to other lengths, the procedure to enlarge $\dH(\cD)$, the examples and
the comparison with known codes.

\section{Preliminaries}\label{sec:prelim}

Throughout, $q$ is a prime power and $m\ge 2$. For $\mathbf{x}=(x_1,\dots,x_n)\in\F_{q^m}^n$
and an integer $k\ge 0$ we write $\mathbf{x}^{q^k}=(x_1^{q^k},\dots,x_n^{q^k})$, and for a set
$V\subseteq \F_{q^m}^n$ we write $V^{q^k}=\{\mathbf{x}^{q^k}:\mathbf{x}\in V\}$. The Hamming weight is
$\wtH(\mathbf{x})=\card{\{i:x_i\neq 0\}}$, where $\card{}$ means cardinality, and for a code $C$ we write $\dH(C)$ for its minimum
Hamming distance, with the convention $\dH(\{0\})=\infty$. If $C\subseteq\F_{q^m}^n$
and $\F_{q^t}\subseteq\F_{q^m}$, the \emph{subfield subcode} of $C$ over $\F_{q^t}$ is
$\subfield{C}{q^t}:=C\cap\F_{q^t}^n$, and the \emph{trace code} is $\Tr_{\F_{q^m}/\F_{q^t}}(C)$,
the trace map from $\F_{q^m}$  to $\F_{q^t}$ being applied componentwise.

We will use Delsarte's theorem~\cite{delsarte}:
\begin{equation}\label{eq:delsarte}
\left(\subfield{C}{q^t}\right)^{\perp}=\Tr_{\F_{q^m}/\F_{q^t}}\left(C^{\perp}\right).
\end{equation}
On $\F_{q^2}^n$ we consider the Hermitian inner product
$\langle \mathbf{x}, \mathbf{y}\rangle_h=\sum_{i}x_iy_i^{q}$, and $C^{\perp_h}$ denotes the Hermitian dual of $C$.
Since $\mathbf{y}\mapsto \mathbf{y}^q$ is an involution of $\F_{q^2}^n$, we have $C^{\perp_h}=(C^{\perp})^{q}$.

Let $q$ be as above, let $p$ be its characteristic, fix $\zeta = \exp(2\pi i/p)$ and  set $\mathrm{tr}:=\mathrm{Tr}_{\F_q / \F_p}$. To measure the performance of a quantum code, one needs a discrete error
model.  For a single $q$-dimensional quantum system $\mathbb{C}^q$, the two
fundamental unitary error operators indexed by elements $a,b\in\F_q$ are
\[
  X(a)\,|x\rangle \;=\; |x+a\rangle,
  \qquad
  Z(b)\,|x\rangle \;=\; \zeta^{\mathrm{tr}(bx)}\,|x\rangle.
\]
$X(a)$ shifts the computational basis by $a$ (\emph{qudit-flip} error) and
$Z(b)$ multiplies each basis vector by a root of unity (\emph{phase-shift}
error).

A \emph{stabilizer code} is a nonzero common eigenspace of commuting tensor products
of the operators $X(a)Z(b)$~\cite{kkks}. The weight of an error is the number of
qudits on which it acts nontrivially. A stabilizer code $Q$ of dimension $q^k$
has parameters $[[n,k,d]]_q$ if it detects all errors of weight less than $d$
but cannot detect some error of weight $d$. It encodes $k$ logical qudits into
$n$ physical qudits and corrects all errors of weight at most
$\lfloor(d-1)/2\rfloor$. The additive-code description for general stabilizer
codes is recalled in Section~\ref{sec:channel}.

The \emph{Hermitian construction}~\cite{kkks} states that if $C\subseteq\F_{q^2}^n$ is an $[n,k]$ linear
code with $C^{\perp_h}\subseteq C$, then there is an $[[n,2k-n,d]]_q$ stabilizer quantum code,
where the minimum distance $d$ is the minimum Hamming weight of a codeword in
$C\setminus C^{\perp_h}$.
The \emph{CSS construction}~\cite{calderbankgoodquantum,css2} states that if $C_2\subseteq C_1\subseteq\F_q^n$ are linear
codes with parameters $[n,k_2]_q$ and $[n,k_1]_q$, then there is an
$[[n,k_1-k_2,\min\{d_X,d_Z\}]]_q$ stabilizer code, where $d_Z = \mathrm{wt}_H(C_2^\perp \setminus C_1^\perp)$ and $d_X = \mathrm{wt}_H(C_1 \setminus C_2)$. As said in the introduction, asymmetric CSS codes have separate $Z$- and
$X$-type distances $d_Z$ and $d_X$, written $[[n,k_1-k_2,d_Z/d_X]]_q$.

The CSS construction with $d_Z\gg d_X$ is the most common approach~\cite{asymmetricConstructions,WFLX,EJLP}.
However, the CSS framework treats the $X$-part and the $Z$-part of an error separately, while it is more natural
to distinguish only two types of errors: diagonal errors and non-diagonal ones. For this purpose, the Hermitian context is more suitable and fits with
the $\lambda$-subfield approach of~\cite{GHW}. It works directly with field elements: an error
operator $X(a)Z(b)$ on a qudit, $a, b \in \F_q$, is identified (after a harmless rescaling explained in Section~\ref{sec:channel}) with the element $\varpi^{1-q}a+b$ of $\F_{q^2}$, where $\{\varpi,\varpi^q\}$ is a normal basis of $\F_{q^2}$ over $\F_q$. Since $\varpi^{1-q}\notin\F_q$, phase-shift errors correspond
to the subfield $\F_q$, and the remaining errors to $\F_{q^2}\setminus\F_q$. This gives a  \emph{single} metric that simultaneously governs both error types. To perform this study, as in~\cite{GHW}, we consider the more general setting of fields $\mathbb{F}_{q^m}$, with $m > 1$.

Let us recall some ideas and definitions from~\cite{GHW}. Fix a parameter $\lambda\ge\tfrac{1}{2}$.
The \emph{$\lambda$-subfield weight} of a single element
$x\in\mathbb{F}_{q^m}$ is defined as

\[
  \wtl(x) :=
  \begin{cases}
    0      & \text{if } x = 0,\\
    1      & \text{if } x \in \mathbb{F}_q\setminus\{0\},\\
    \lambda & \text{if } x \in \mathbb{F}_{q^m}\setminus\mathbb{F}_q.
  \end{cases}
\]

The weight is extended additively and coordinate-wise to vectors
$\mathbf{x}=(x_1,\dots,x_n)\in\mathbb{F}_{q^m}^n$.
The corresponding \emph{$\lambda$-subfield distance} is
$\dl(\mathbf{x},\mathbf{y}):=\wtl(\mathbf{x}-\mathbf{y})$,
which defines a genuine metric on $\mathbb{F}_{q^m}^n$.
Note that $\lambda=1$ recovers the ordinary Hamming metric.

\begin{remark}\label{rem:lee}
The joint quantum Lee weight studied in~\cite{GuoEtAlLee}
assigns weight $1$ to $X$ and $Z$, and weight $2$ to $Y$.
Consequently, for $q=2$ and $m=2$, it coincides with twice
the $\tfrac12$-subfield weight after exchanging $Y$ and $Z$
in the identification of Pauli errors with $\F_4$.
This exchange is implemented, up to phases, by a single-qubit
Clifford gate, and the equality extends additively to Pauli
strings. Thus, the binary joint Lee metric corresponds to
the boundary case $\lambda=\tfrac12$, up to a Clifford
relabeling and scaling. Our main bounds and constructions
concern the regime $\lambda\ge1$.
\end{remark}

For a linear code $C \subseteq\mathbb{F}_{q^m}^n$
the \emph{minimum $\lambda$-subfield distance} is
\[
  d_\lambda(C) :=
  \min\bigl\{d_\lambda(\mathbf{x},\mathbf{y}) \;:\; \mathbf{x},\mathbf{y}\in C,\; \mathbf{x}\neq \mathbf{y}\bigr\}.
\]
A code with minimum $\lambda$-subfield distance $d$ can uniquely correct
every error vector $e$ with $\mathrm{wt}_\lambda(e)<d/2$ (note that $d$ and the $\lambda$-weights need not be integers).
More generally, as for the Hamming metric, such a code corrects every error of a set $E_1$ (with $0\in E_1$) while detecting every error of a set $E_2$ provided that $\wtl(e_1-e)<d$ for all $e_1\in E_1$ and $e\in E_1\cup E_2$ with $e\neq e_1$.
The central role of $d_\lambda(C)$ in this framework mirrors
that of the minimum Hamming distance in classical coding theory: it
controls the error-correction capability of the code against the
\emph{asymmetric} error model in which errors from
$\mathbb{F}_{q^m}\setminus\mathbb{F}_q$ are penalized differently by a factor $\lambda$, and hence more heavily when $\lambda>1$, than errors confined to the base field.
The larger $\lambda$ is, the more important it becomes to keep
the number of off-base-field error positions small.

An entry $x_i$ of a vector $\mathbf{x}\in\mathbb{F}_{q^m}^n$ is said to be of
\emph{base type} if $x_i\in\mathbb{F}_q^*$ and of \emph{roof type}
if $x_i\in\mathbb{F}_{q^m}\setminus\mathbb{F}_q$.
The $\lambda$-subfield weight is entirely determined by the number
of entries of each type, but it conflates these two pieces of
information into a single scalar.
To retain the full detail, we can consider the
\emph{base weight} and the \emph{roof weight}:
\[
  \mathrm{wt}_b(\mathbf{x}) := \card\{i \;:\; x_i\in\mathbb{F}_q\setminus\{0\}\},
  \qquad
  \mathrm{wt}_r(\mathbf{x}) := \card\{i \;:\; x_i\in\mathbb{F}_{q^m}\setminus\mathbb{F}_q\}.
\]
Their joint encoding is the \emph{base-roof (BR)-weight}
$\mathrm{wt}_{BR}(\mathbf{x}):=(\mathrm{wt}_b(\mathbf{x}),\,\mathrm{wt}_r(\mathbf{x}))$,
a vector in $\mathbb{N}_0^2$.
The BR-weight completely determines the $\lambda$-subfield weight: if
$\mathrm{wt}_{BR}(\mathbf{x})=(s,t)$, then
$\mathrm{wt}_\lambda(\mathbf{x})=s+\lambda t$ for every $\lambda$.
In the same way one defines the \emph{base distance}
$d_b(\mathbf{x},\mathbf{y}):=\mathrm{wt}_b(\mathbf{x}-\mathbf{y})$ and the
\emph{roof distance} $d_r(\mathbf{x},\mathbf{y}):=\mathrm{wt}_r(\mathbf{x}-\mathbf{y})$; neither
alone is a true metric on $\mathbb{F}_{q^m}^n$, but the roof
distance satisfies the triangle inequality (making it a
pseudometric), and their pair forms the
\emph{BR-distance} $d_{BR}(\mathbf{x},\mathbf{y}):=(d_b(\mathbf{x},\mathbf{y}),d_r(\mathbf{x},\mathbf{y}))$.

For a code $C \subseteq\mathbb{F}_{q^m}^n$,
the \emph{set of BR-minimal distances} is
\[
  d_{BR}(C) :=
  \min\bigl\{d_{BR}(\mathbf{x},\mathbf{y})\in\mathbb{N}_0^2 \;:\; \mathbf{x},\mathbf{y}\in C,\;\mathbf{x}\neq \mathbf{y}\bigr\},
\]
where the minimum is taken with respect to the componentwise
partial order $(s,t)\preceq(s',t')\Leftrightarrow s\le s'$
and $t\le t'$, and in general $d_{BR}(C)$ consists of
several minimal pairs.
The {\it minimum $\lambda$-subfield distance of $C$} is then recovered as
\[
  d_\lambda(C) = \min\bigl\{s+\lambda t \;:\; (s,t)\in d_{BR}(C)\bigr\}.
\]
Since the BR-weight carries strictly more information than the
$\lambda$-subfield weight alone, working with BR-distances makes it
possible to describe more precisely which combinations of base-type
and roof-type errors a code can correct.

\section{A lower bound for the subfield distance of a linear code}\label{sec:bound}

In this section we establish a lower bound for the subfield distance of a linear code. The bound is stated in terms of the minimum distances of certain associated codes, and is the key tool for the constructions in later sections. We divide our study into three subsections.

\subsection{A lower bound for the subfield distance}\label{subsec:lowerbound}

\begin{definition}\label{def:D}
Let $C\subseteq \F_{q^m}^n$ be a linear code. We define
\[
  \cD=\cD(C):=\{c-c^q : c\in C\}\subseteq\F_{q^m}^n.
\]
\end{definition}

\begin{lemma}\label{lem:Dlinear}
$\cD$ is an $\F_q$-linear code. If $C\neq\{0\}$, then $\cD\neq\{0\}$.
\end{lemma}
\begin{proof}
Let $a,b\in\F_q$ and $c_1,c_2\in C$. Since $a^q=a$ and $b^q=b$,
\[
  a(c_1-c_1^q)+b(c_2-c_2^q)=(ac_1+bc_2)-(ac_1+bc_2)^q\in\cD.
\]
Now let $0\neq c\in C$. If $c\notin\F_q^n$, then $c-c^q\neq 0$. Otherwise take
$\alpha\in\F_{q^m}\setminus\F_q$ (recall $m\ge2$); every nonzero entry $\alpha c_i$ of
$\alpha c\in C$ lies outside $\F_q$; thus $\alpha c-(\alpha c)^q\neq 0$.
\end{proof}

\begin{remark}
$\cD$ is not $\F_{q^m}$-linear in general. For instance, if $n=1$ and $C=\F_{q^m}$, then
$\cD$ is the image of the $\F_q$-linear map $x\mapsto x-x^q$, whose kernel is $\F_q$; hence
$\dim_{\F_q}\cD=m-1$, and $\cD$ is neither $\{0\}$ nor $\F_{q^m}$, the only
$\F_{q^m}$-subspaces of $\F_{q^m}$.
\end{remark}

\begin{definition}
We denote by $\bcD:=\operatorname{Span}_{\F_{q^m}}(\cD)$ the $\F_{q^m}$-linear code spanned
by $\cD$. Note that $\bcD\subseteq C+C^q$. For convenience we will mostly work with $\cD$,
but all the bounds below remain valid if $\cD$ is replaced by $\bcD$, which is often easier to
estimate.
\end{definition}

Since $\cD\subseteq\bcD$, we clearly have the following inequality.

\begin{lemma}\label{lem:DbarD}
$\dH(\cD)\ge\dH(\bcD)$.
\end{lemma}

Let us state a result involving the codes $C$, $\cD$ and $\bcD$.

\begin{lemma}\label{lem:r}
Let $C\subseteq\F_{q^m}^n$ be a nonzero linear code and let $c\in C$ with BR-weight
$\wt_{BR}(c)=(b,r)$. Then, the following equality and inequalities hold:
\begin{enumerate}[label=(\alph*)]
\item $\wtH(c-c^q)=r$.
\item If $r\neq0$, then $r\ge\dH(\cD)\ge\dH(\bcD)$.
\item $\dH(C)\ge\dH(\cD)\ge\dH(\bcD)$.
\end{enumerate}
\end{lemma}
\begin{proof}
(a) For $x\in\F_{q^m}$ we have $x-x^q=0$ if and only if $x\in\F_q$.

(b) If $r\neq0$, then $c-c^q$ is a nonzero element of $\cD$ of weight $r$ by (a); now use
Lemma~\ref{lem:DbarD}.

(c) Let $c\in C$ be a codeword of weight $\dH(C)$, with $\wt_{BR}(c)=(b,r)$. If $r\neq 0$,
then $\dH(C)=b+r\ge r\ge\dH(\cD)$ by (b). If $r=0$, then $c\in\F_q^n$, and for
$\alpha\in\F_{q^m}\setminus\F_q$ the codeword $\alpha c$ has BR-weight $(0,\dH(C))$; therefore $\dH(C)\ge\dH(\cD)$ again by (b).
\end{proof}

Our main result in this section provides a lower bound for the minimum $\lambda$-subfield distance of a nonzero linear code.

\begin{theorem}\label{thm:main}
Let $\lambda\ge1$ and let $C\subseteq\F_{q^m}^n$ be a nonzero linear code. Then
\[
  \dl(C)\ge\min\left\{(\lambda-1)\dH(\cD)+\dH(C),\ \dH(\subfield{C}{q})\right\},
\]
where, if $\subfield{C}{q}=\{0\}$, the minimum is just $(\lambda-1)\dH(\cD)+\dH(C)$ (recall that
$\dH(\{0\})=\infty$). The same result holds with $\cD$ replaced by $\bcD$.
\end{theorem}
\begin{proof}
Let $0\neq c\in C$ with $\wt_{BR}(c)=(b,r)$. If $r\neq 0$, since $b+r=\wtH(c)\ge\dH(C)$,
$\lambda-1\ge0$ and $r\ge\dH(\cD)$ by Lemma~\ref{lem:r}, it holds
\[
  \wtl(c)=b+\lambda r=(b+r)+(\lambda-1)r\ge \dH(C)+(\lambda-1)\dH(\cD).
\]
If $r=0$, then $0\ne c\in\subfield{C}{q}$ and $\wtl(c)=b=\wtH(c)\ge\dH(\subfield{C}{q})$.

The last statement follows from Lemma~\ref{lem:DbarD}.
\end{proof}

\begin{remark}
For $1/2\le\lambda<1$ we have $\wtl(c)=b+\lambda r\ge\lambda(b+r)=\lambda\wtH(c)$ for every $c\in C$, hence $\dl(C)\ge\lambda\dH(C)$.
\end{remark}

For large $\lambda$ the $\lambda$-subfield distance is the Hamming distance of the subfield subcode. Let us state the meaning of a large $\lambda$, which follows directly from Theorem~\ref{thm:main}.

\begin{corollary}\label{cor:large}
Assume $\subfield{C}{q}\neq\{0\}$. Then $\dl(C)\le\dH(\subfield{C}{q})$ for every $\lambda>0$, and
\[
  \dl(C)=\dH(\subfield{C}{q})\quad\text{for every }\lambda\ge\frac{\dH(\subfield{C}{q})-\dH(C)}{\dH(\cD)}+1.
\]
\end{corollary}
%\begin{proof}
%A codeword of $\subfield{C}{q}$ of weight $\dH(\subfield{C}{q})$ has $\lambda$-weight %$\dH(\subfield{C}{q})$,
%which gives the upper bound. If $\lambda$ satisfies the stated inequality, then $\lambda\ge1$
%(as $\dH(\subfield{C}{q})\ge\dH(C)$) and $(\lambda-1)\dH(\cD)+\dH(C)\ge\dH(\subfield{C}{q})$, so
%Theorem~\ref{thm:main} gives $\dl(C)\ge\dH(\subfield{C}{q})$.
%\end{proof}
The Hermitian construction makes the case $m=2$ particularly interesting for quantum error correction.

\begin{lemma}\label{lem:dimCq}
Let $m=2$ and let $C\subseteq\F_{q^2}^n$ be a linear code of dimension $k$. Then,
$$\dim_{\F_q}\subfield{C}{q}=\dim_{\F_{q^2}}(C\cap C^q)\ge 2k-n.$$
\end{lemma}
\begin{proof}
$V=C\cap C^q$ satisfies $V^q=V$. Let $\varpi\in\F_{q^2}\setminus\F_q$. For $v\in V$, the
vectors $v+v^q$ and $\varpi v+(\varpi v)^q$ lie in $V\cap\F_q^n$, and
$\varpi(v+v^q)-(\varpi v+(\varpi v)^q)=(\varpi-\varpi^q)v^q$. Hence $v^q$ (and, applying this
to $v^q\in V$, also $v$) lies in the $\F_{q^2}$-span of $V\cap\F_q^n$. Since
$\F_q$-independent vectors of $\F_q^n$ are $\F_{q^2}$-independent, we get
$\dim_{\F_q}(V\cap\F_q^n)=\dim_{\F_{q^2}} V$. Finally $V\cap\F_q^n=\subfield{C}{q}$, and
$$
\dim_{\F_{q^2}} V=\dim_{\F_{q^2}}  C+\dim_{\F_{q^2}}  C^q-\dim_{\F_{q^2}} (C+C^q)\ge 2k-n.
$$
\end{proof}

\begin{remark}\label{rem:quantumbound}
By Lemma~\ref{lem:dimCq}, the quantum code $[[n,2k-n]]_q$ obtained from a Hermitian
dual-containing code $C\subseteq\F_{q^2}^n$ satisfies $2k-n\le\dim_{\F_q}\subfield{C}{q}$. As
we have seen, $\dH(\subfield{C}{q})$ is the quantity that governs $\dl(C)$ for large $\lambda$;
thus the dimension of the quantum code is bounded by the dimension of a $q$-ary code whose
minimum distance is equal to the asymptotic value of $\dl(C)$.
\end{remark}

\subsection{The trace code and the exact value of the subfield distance}\label{subsec:trace}

In the following, we will show the connection between $\mathcal{D}$ and the trace code of $C$. We set $\Tr=\Tr_{\mathbb{F}_{q^m}/\mathbb{F}_q}$, and for $c\in\mathbb{F}_{q^m}^n$, we write
\begin{equation}
\label{laR}
r(c)=\#\{i : c_i\notin\mathbb{F}_q\}.
\end{equation}
Note that $\mathrm{wt}_{BR}(c)=(\mathrm{wt}_H(c)-r(c), r(c))$.

\begin{proposition}\label{prop:trace}
Let $C\subseteq\F_{q^m}^n$ be a linear code.
\begin{enumerate}[label=(\alph*)]
\item For every $c\in C\setminus\subfield{C}{q}$, there is a nonzero $v\in\Tr(C)$ with
$\operatorname{supp}(v)\subseteq\{i:c_i\notin\F_q\}$. In particular $r(c)\ge\dH(\Tr(C))$ and
$\dH(\cD)\ge\dH(\Tr(C))$. 
\item If $m=2$ and $\varpi\in\F_{q^2}\setminus\F_q$, then $\cD=(\varpi-\varpi^q)\Tr(C)$ and
$\bcD=(\varpi-\varpi^q)\big(\Tr(C)\otimes\F_{q^2}\big)$. Hence $\dH(\cD)=\dH(\bcD)=\dH(\Tr(C))$, and
$\dim_{\F_q}\cD=\dim_{\F_q}\Tr(C)=2\dim C-\dim_{\F_q}\subfield{C}{q}$.
\end{enumerate}
\end{proposition}
\begin{proof}
We begin by proving (a). Let $K_0=\{\theta\in\F_{q^m}:\Tr(\theta)=0\}$, an $\F_q$-subspace of dimension $m-1$. Since the trace form
$(x,y)\mapsto\Tr(xy)$ is nondegenerate, the orthogonal of $K_0$ with respect to the trace form is one-dimensional, and it contains $1$; hence
$x\in\F_q$ if and only if $\Tr(\theta x)=0$ for all $\theta\in K_0$. Let $c\in C\setminus\subfield{C}{q}$ and choose $i$
with $c_i\notin\F_q$ and $\theta\in K_0$ with $\Tr(\theta c_i)\neq0$. Then $v=\Tr(\theta c)\in\Tr(C)$ because
$\theta c\in C$. Note that $v\neq0$ and $v_j=c_j\Tr(\theta)=0$ whenever $c_j\in\F_q$. Thus
$r(c)\ge\wtH(v)\ge\dH(\Tr(C))$, and the last claim follows from Lemma~\ref{lem:r}(a).

Now, let us see (b). Set $\delta_0=\varpi-\varpi^q\neq0$; then $\delta_0^q=-\delta_0$, and $\psi(x):=(x-x^q)/\delta_0$ satisfies
$\psi(x)^q=\psi(x)$. Then $\psi$ is a nonzero $\F_q$-linear map $\F_{q^2}\to\F_q$, and therefore
$\psi(x)=\Tr(\theta_0x)$ for some $\theta_0\neq0$. Hence $c-c^q=\delta_0\Tr(\theta_0c)$ and, since $\theta_0C=C$, we have $\cD=\delta_0\Tr(C)$. The $\F_{q^2}$-span of $\Tr(C)$ is $\Tr(C)\otimes\F_{q^2}$ (an $\F_q$-basis of $\Tr(C)$ is
$\F_{q^2}$-independent), and a nonzero vector $u+\varpi v$ of it, with $u,v\in\Tr(C)$, has weight at least
$\max\{\wtH(u),\wtH(v)\}\ge\dH(\Tr(C))$. Finally, the kernel of the $\F_q$-linear map $c\mapsto c-c^q$ on $C$ is
$\subfield{C}{q}$, which concludes the proof.
\end{proof}

Consequently Theorem~\ref{thm:main} implies
\begin{equation}\label{eq:maintrace}
\dl(C)\ge\min\{(\lambda-1)\dH(\Tr(C))+\dH(C),\ \dH(\subfield{C}{q})\},
\end{equation}
where, by Delsarte's theorem~\eqref{eq:delsarte}, $\Tr(C)=(C^\perp\cap\F_q^n)^\perp$; for $m=2$ it coincides, by Proposition~\ref{prop:trace}(b), with the
bound of Theorem~\ref{thm:main}.

\begin{remark}\label{rem:grs}
Let $C=\operatorname{GRS}_k(\mathbf{a},\mathbf{v})\subseteq\F_{q^2}^n$ be a \textit{generalized Reed-Solomon code}, where the entries of
$\mathbf{a}$ are distinct, the entries of $\mathbf{v}$ are nonzero, and
$\operatorname{GRS}_k(\mathbf{a},\mathbf{v})=\{(v_i f(a_i))_{i=1}^n:\deg f<k\}$.
Since $\dH(C)=n-k+1$, Inequality~\eqref{eq:maintrace} becomes
\[
  \dl(C)\ge\min\{n-k+1+(\lambda-1)\dH(\Tr(C)),\ \dH(\subfield{C}{q})\},
  \qquad \lambda\ge1.
\]
Here $\subfield{C}{q}$ is an alternant code. Its usual designed-distance estimate
$\dH(\subfield{C}{q})\ge n-k+1$ alone yields no improvement over $\dH(C)=n-k+1$;
an improvement requires a stronger estimate for this subfield subcode. If both
$\mathbf{a}$ and $\mathbf{v}$ have entries in $\F_q$, then $\subfield{C}{q}$ is the
corresponding GRS code over $\F_q$, so $\dl(C)=n-k+1$ for every $\lambda\ge1$.

Extension-field multipliers can nevertheless give a strict improvement. For example, let
$\F_9=\F_3(\mu)$ with $\mu^2=-1$, and consider
\[
  h=(1,\mu,1+\mu,1-\mu),\qquad
  C=\{c\in\F_9^4:h\cdot c=0\}
   =\operatorname{GRS}_3\big((0,1,\mu,-\mu),(-1,\mu,1,1)\big).
\]
Then $C$ is a $[4,3,2]_9$ code and $C^{\perp_h}=\langle h^3\rangle\subseteq C$, since
$\sum_i h_i^4=0$. The subfield subcode has parity-check matrix
\[
  \begin{pmatrix}1&0&1&1\\0&1&1&-1\end{pmatrix},
\]
and hence has parameters $[4,2,3]_3$. Proposition~\ref{prop:trace}(b) gives
$\dim_{\F_3}\Tr(C)=4$, so $\Tr(C)=\F_3^4$ and $\dH(\cD)=1$. Thus
\(
  \dl(C)=\min\{\lambda+1,3\},
\)
for $\lambda \geq 1$. The lower bound follows from Theorem~\ref{thm:main}, and the two upper bounds follow from
$(1,\mu,0,0)\in C$ and a weight-three subfield codeword. The resulting $[[4,2]]_3$
quantum code has the same $\lambda$-subfield distance, since every nonzero word of
$C^{\perp_h}$ has Hamming weight four.
\end{remark}

For Hermitian dual-containing codes there is an unavoidable trade-off between the two terms
of~\eqref{eq:maintrace}.

\begin{proposition}\label{prop:tradeoff}
Let $C\subseteq\F_{q^2}^n$ be linear with $C^{\perp_h}\subseteq C$. Then $(\subfield{C}{q})^{\perp}\subseteq\Tr(C)$.
Consequently $\dH(\cD)\le\dH((\subfield{C}{q})^\perp)$ and $\dim_{\F_q}\cD\ge n-\dim_{\F_q}\subfield{C}{q}$.
\end{proposition}
\begin{proof}
By~\eqref{eq:delsarte}, $(\subfield{C}{q})^\perp=\Tr(C^\perp)$, and $C^\perp=(C^{\perp_h})^q\subseteq C^q$. Since
$\Tr(c^q)=\Tr(c)$, we get $(\subfield{C}{q})^\perp\subseteq\Tr(C^q)=\Tr(C)$. Finally, the result follows from Proposition~\ref{prop:trace}(b).
\end{proof}

Thus a subfield subcode $\subfield{C}{q}$ of large minimum distance (which is what the second term of the bound needs) has to be small, which forces $\Tr(C)$ to contain the large code $(\subfield{C}{q})^\perp$ and $\dH(\cD)$ to be small; on the other hand, the quantum dimension $k_Q\le\dim_{\F_q}\subfield{C}{q}$ (Remark~\ref{rem:quantumbound}) pushes in the opposite direction on $\dim_{\F_q}\subfield{C}{q}$.
This explains why, in the examples of the forthcoming Section~\ref{sec:hermitian}, the first term within the minimum of~\eqref{eq:maintrace} grows slowly with $\lambda$. The following exact expression shows what is lost.

\begin{proposition}\label{prop:profile}
Let $C\subseteq\F_{q^m}^n$ be a nonzero linear code and, for $r\ge1$, let
$$W_r(C):=\min\{\wtH(c):c\in C,\ r(c)=r\}\; (\mbox{with } \; \min\emptyset=\infty).$$ Then, for every $\lambda\ge1$,
\[
\dl(C)=\min\Big\{\dH(\subfield{C}{q}),\ \min_{r\ge1}\big(W_r(C)+(\lambda-1)r\big)\Big\}
=\min_{0\neq c\in C}\big(\lambda\wtH(c)-(\lambda-1)M(c)\big),
\]
where $M(c)$ is the largest number of nonzero coordinates of $c$ lying in a common class $x\F_q^*$,
$x\in\F_{q^m}^*$. Moreover, $$W_r(C)\ge\max\{\dH(C),r\},\; \mbox{ and } \; W_r(C)=\infty \mbox{ if } r<\dH(\cD).$$ In particular
$\lambda\mapsto\dl(C)$ is a nondecreasing concave piecewise linear function on $[1,\infty)$.
\end{proposition}
\begin{proof}
The first equality follows from $\wtl(c)=\wtH(c)+(\lambda-1)r(c)$. For the second, note that for $\xi\in\F_{q^m}^*$
the codeword $\xi c$ has $\mathrm{wt}_b(\xi c)$ equal to the number of coordinates of $c$ in the class $\xi^{-1}\F_q^*$, so that
$\min_\xi\wtl(\xi c)=\lambda\wtH(c)-(\lambda-1)M(c)$. The inequalities for $W_r$ are clear, and the last statement
holds because $\dl(C)$ is a minimum of finitely many nondecreasing affine functions of $\lambda$.
\end{proof}

Theorem~\ref{thm:main} is the result of replacing each $W_r(C)$ by $\dH(C)$ and $r$ by $\dH(\cD)$. For a cyclic
code we may assume that $0\in\{i:c_i\notin\F_q\}$, so $W_1(C)$ and $W_2(C)$ can be bounded from below with $1$ and $n-1$
minimum distance computations of small $\F_q$-linear codes, namely
\[
E_0=\{c\in C:c_i\in\F_q \text{ for }i\neq0\},\qquad
E_j=\{c\in C:c_i\in\F_q \text{ for }i\notin\{0,j\}\},
\quad 1\le j\le n-1.
\]
Indeed,
\(W_1(C)\ge\dH(E_0),\) and $W_2(C)\ge\min_{1\le j\le n-1}\dH(E_j)$.
Together with $W_r(C)\ge\max\{\dH(C),r\}$ for $r\ge3$, these estimates can be used in Proposition~\ref{prop:profile}
to improve the bound considerably in some of the examples below.

Let us see that the $\lambda$-subfield distance can also be computed exactly with standard tools for the Hamming metric.  The idea is to replace each coordinate by a block over $\F_q$ whose Hamming weight is a fixed multiple of the $\lambda$-subfield weight of the original coordinate.

Let $\lambda \geq 1$ be a rational number and consider integers $A\ge0$ and $B\ge1$ such that $\lambda=1+A/(qB)$. Let $\Theta$ be a set of representatives of
$\F_{q^m}^*/\F_q^*$ and $\Theta_0=\{\theta\in\Theta:\Tr(\theta)=0\}$. Define  $$\varphi:\F_{q^m}\to\F_q^{N},$$ where
$N=B\,\card\Theta+A\,\card\Theta_0$, as the map  that sends $x$ to the vector formed by $B$ copies of $(\Tr(\theta x))_{\theta\in\Theta}$
and $A$ copies of $(\Tr(\theta x))_{\theta\in\Theta_0}$.

\begin{proposition}\label{prop:image}
With the above notation it holds that  $\wtH(\varphi(x))=Bq^{m-1}\wtl(x)$ for all
$x\in\F_{q^m}$. Consequently, if $C\subseteq\F_{q^m}^n$ is linear of dimension $k$, then $\varphi(C)$ (applied
coordinatewise) is an $[nN,mk]_q$ linear code, of length $nN$ over $\F_q$, and
\[
\dl(C)=\frac{\dH(\varphi(C))}{Bq^{m-1}} .
\]
\end{proposition}
\begin{proof}
Let $x \in \F_{q^m}^*$. Consider the nonzero $\F_q$-linear map $\F_{q^m} \rightarrow \F_{q}$, $y\mapsto\Tr(y x)$. When $\theta\in\Theta$,  $\Tr(\theta x)\neq0$ for exactly
$(q^m-q^{m-1})/(q-1)=q^{m-1}$ elements of $\Theta$, and, among the elements of $\Theta_0$, for none if
$x\in\F_q^*$ and for $(q^{m-1}-q^{m-2})/(q-1)=q^{m-2}$ if $x\notin\F_q$. Therefore, $\wtH(\varphi(x))$ is $Bq^{m-1}$ for
$x\in\F_q^*$ and $Bq^{m-1}+Aq^{m-2}=Bq^{m-1}\lambda$ for $x\notin\F_q$, which proves our first statement. Finally, the map $\varphi$ is $\F_q$-linear and
injective, which gives the dimension.
\end{proof}

\begin{remark}
Proposition~\ref{prop:image} allows one to compute $\dl(C)$ with the Brouwer--Zimmermann algorithm~\cite{Grassl06}. We have used it, with
Magma~\cite{magma}, to check exact values in the forthcoming Section~\ref{sec:hermitian}. Proposition~\ref{prop:image} also permits us to transfer every upper bound for
$q$-ary linear codes (Griesmer, linear programming, tables) to $\dl$.
\end{remark}

\subsection{Decoding up to half the lower bound}\label{subsec:decoding}

The lower bound of Theorem~\ref{thm:main} also admits a decoding interpretation.
The following reduction attains half of our lower bound using three decoders for the Hamming metric. If a code is $\{0\}$, we use the decoder that always returns $0$. 

\begin{proposition}\label{prop:decoding}
Let $C\subseteq\F_{q^m}^n$ be a linear code, and let
$\lambda\ge1$. Suppose that $\delta_C,\delta_{\cD},\delta_q$ are positive integers (with $\delta_q=\infty$ when $\subfield{C}{q}=\{0\}$) lower
bounds for $\dH(C),\dH(\cD),\dH(\subfield{C}{q})$, respectively, and that Hamming decoders are available
for these three codes which correct every error of weight strictly less than half the
corresponding bound. Set
\begin{equation}\label{eq:decodingbound}
  L=\min\{\delta_C+(\lambda-1)\delta_{\cD},\delta_q\}.
\end{equation}
Then every error $e$ with $\wtl(e)<L/2$ can be corrected using at most one call to each
of the three decoders, and linear algebra over $\F_q$.
In particular, the complexity of the resulting algorithm is polynomial if the three
component decoders are.
\end{proposition}
\begin{proof}
Write $F(x)=x-x^q$, applied coordinatewise, so that $F(C)=\cD$ and
$\ker(F|_C)=\subfield{C}{q}$. Fix an $\F_q$-linear projection $\pi:\F_{q^m}\to\F_q$ which is the
identity on $\F_q$; for example, take the first coordinate in a basis beginning with $1$.
We also apply $\pi$ coordinatewise. Given a received word $y$, proceed as follows.
\begin{enumerate}[label=(\roman*)]
\item Apply the Hamming decoder for $C$ to $y$. If it returns $\widehat c\in C$ with
$\wtl(y-\widehat c)<L/2$, output $\widehat c$.
\item Otherwise, apply the Hamming decoder for $\cD$ to $F(y)$. If it returns
$d_0\in\cD$, find $c_0\in C$ such that $F(c_0)=d_0$ by solving an $\F_q$-linear system.
\item Apply the Hamming decoder for $\subfield{C}{q}$ to $\pi(y-c_0)$. If it returns $a\in\subfield{C}{q}$ with
$\wtl(y-c_0-a)<L/2$, output $c_0+a$. If a required decoding or membership test fails,
or no candidate passes the final weight test, report failure.
\end{enumerate}
By Theorem~\ref{thm:main}, $\dl(C)\ge L$, so at most one codeword can pass the weight
test. Thus it remains to prove that one of the two routes succeeds when $y=c+e$,
$c\in C$, and $\wtl(e)<L/2$.

Let $\wt_{BR}(e)=(b,t)$ and put $h=b+t=\wtH(e)$. If $2h<\delta_C$, step~(i)
returns $c$. Otherwise $2h\ge\delta_C$. This case is impossible when $\lambda=1$,
because $2h=2\wtl(e)<L\le\delta_C$. When $\lambda>1$, we have
\[
  2h+2(\lambda-1)t=2\wtl(e)<L
  \le\delta_C+(\lambda-1)\delta_{\cD},
\]
which forces $2t<\delta_{\cD}$. Since $\wtH(F(e))=t$, step~(ii) recovers $F(c)$.
Consequently $F(c-c_0)=d_0-d_0=0$, i.e., we have $c-c_0\in\subfield{C}{q}$ and
\[
  \pi(y-c_0)=(c-c_0)+\pi(e),\qquad
  \wtH(\pi(e))\le h\le\wtl(e)<L/2\le\delta_q/2.
\]
The decoder for $\subfield{C}{q}$ therefore returns $a=c-c_0$, and step~(iii) outputs $c$.
The projection and a linear right inverse of $F|_C:C\to\cD$ can be precomputed
by linear algebra over $\F_q$, proving the complexity assertion.
\end{proof}

If $m=2$ and $C^{\perp_h}\subseteq C$, the same procedure can be used for the
associated quantum code. A measured stabilizer syndrome specifies a coset $e+C$. Choose
any representative $y=c+e$, with $c\in C$, and Proposition~\ref{prop:decoding} recovers $e=y-c$ whenever
$\wtl(e)<L/2$.
For cyclic codes, suitable component decoders are described at the end of
Section~\ref{sec:cyclic}.

\section{Biased noise and comparison with asymmetric CSS codes}\label{sec:channel}

This section justifies the metric for a biased channel (Proposition~\ref{prop:channel}), relates $\dl$ to $(d_Z,d_X)$ for stabilizer and CSS codes (Proposition~\ref{prop:dzdx}), and proves a comparison theorem with pure asymmetric CSS codes (Theorem~\ref{thm:dominance}).

\subsection{The channel}
Let $X(a)Z(b)$, $a,b\in\F_q$, be the error basis on a qudit~\cite{kkks}; the error is diagonal (a phase-shift error)
if and only if $a=0$. We consider the memoryless channel in which each qudit is affected by a given nontrivial
diagonal error with probability $p_Z/(q-1)$, by a given non-diagonal error with probability $p_X/(q^2-q)$, and by
no error with probability $p_0=1-p_Z-p_X$ (so all $q^2-q$ non-diagonal errors are assumed to be equally likely; for qubits, $X$- and $Y$-errors have the same probability). Let $\{\varpi, \varpi^q\}$ be a normal basis of $\F_{q^2}$ over $\F_q$. Following~\cite{kkks}, we identify an element $(a,b) \in \F_q \times \F_q$ with $a\varpi + b \varpi^q$. Multiplying by the constant $\varpi^{-q}$ (which changes neither $\F_{q^2}$-linear codes nor their Hermitian duals) we identify $X(a)Z(b)$ with $\varpi^{1-q}a+b\in\F_{q^2}$. Since $\varpi^{1-q}\in\F_q$ would give $\varpi^q=c\varpi$ with $c\in\F_q$, contradicting that $\{\varpi,\varpi^q\}$ is a basis, we have $\varpi^{1-q}\notin\F_q$. Thus,  the diagonal errors are the nonzero elements of $\F_q$, and we identify an error on $n$ qudits with a vector $\varepsilon\in\F_{q^2}^n$.

\begin{proposition}\label{prop:channel}
Assume $p_X, p_Z \neq 0$, $p_Z/(q-1)<p_0$ and $p_X/(q^2-q)\le p_Z/(q-1)$, and set
\[
\alpha=\ln\frac{(q-1)p_0}{p_Z},\qquad
\lambda=\frac{1}{\alpha}\ln\frac{(q^2-q)p_0}{p_X}=1+\frac{\ln(qp_Z/p_X)}{\alpha}\ \ge1 .
\]
Then the probability of the error $\varepsilon$ is $p_0^n\exp(-\alpha\wtl(\varepsilon))$. In particular, among the
errors compatible with a syndrome, the most likely ones are those of minimum $\lambda$-subfield weight.
\end{proposition}
\begin{proof}
If $\wt_{BR}(\varepsilon)=(b,r)$, the probability is
$p_0^{n-b-r}(p_Z/(q-1))^b(p_X/(q^2-q))^r=p_0^n e^{-\alpha b}e^{-\alpha\lambda r}$.
\end{proof}

Thus, the $\lambda$-subfield metric plays for this channel the role that the Hamming metric plays for the symmetric
channel. If a quantum code has $\lambda$-subfield distance $d$, every error of $\lambda$-weight less
than $d/2$ is corrected, and every uncorrectable error has probability at most $p_0^ne^{-\alpha d/2}$.
For a fixed code and bounded distance decoder, as $p_Z,p_X\to0$ with $\lambda$ kept fixed
(so the bias $\eta$ grows when $\lambda>1$), the decoding failure probability is asymptotically proportional to
$e^{-\alpha w^*}$, where $w^*\ge d/2$ is the smallest $\lambda$-weight of an error that is not corrected.
Therefore, a natural worst-case figure of merit of a code for this channel is
its $\lambda$-subfield distance, with $\lambda$ determined by the channel probabilities
as in Proposition~\ref{prop:channel}.

\begin{remark}\label{rem:lambdavalues}
For qubits, with bias $\eta=p_Z/p_X$ and $p_X,p_Z\neq 0$ (here $p_X$ is the total probability of $X$- and $Y$-errors), we have $\lambda=1+\ln(2\eta)/\ln(p_0/p_Z)$. Strongly biased noise is the regime targeted, e.g., by stabilized cat qubits~\cite{GM,Puri,Putterman} and by surface codes tailored to biased noise~\cite{TBF,XZZX}. Table~\ref{tab:lambda} shows that realistic
values of $\lambda$ are moderate; for instance, if $99\%$ of the errors are phase flips, $\lambda$ is $2.15$, $2.79$
and $3.40$ for total error probabilities of $1\%$, $5\%$ and $10\%$, respectively. 
Note also that $\lambda\to1$ when $p_Z\to0$ with $\eta$ fixed,
so the $\lambda$-subfield metric approaches the Hamming metric.
For $\lambda\ge(\dH(\subfield{C}{q})-\dH(C))/\dH(\cD)+1$, $\dl(C)$ already
reaches its pure-dephasing limit $\dH(\subfield{C}{q})$
(Corollary~\ref{cor:large}), even though $p_X$ may still be positive.
\end{remark}

\begin{table}[ht]
\centering
\caption{The value of $\lambda$ in Proposition~\ref{prop:channel} for qubits, as a function of the probability $p_Z$
of a phase flip and of the bias $\eta=p_Z/p_X$.}\label{tab:lambda}
\begin{tabular}{c|cccc}
\toprule
$p_Z\ \backslash\ \eta$ & $10$ & $10^2$ & $10^4$ & $10^6$\\
\midrule
$10^{-2}$ & $1.65$ & $2.15$ & $3.16$ & $4.16$\\
$10^{-3}$ & $1.43$ & $1.77$ & $2.43$ & $3.10$\\
$10^{-4}$ & $1.33$ & $1.58$ & $2.08$ & $2.58$\\
\bottomrule
\end{tabular}
\end{table}

\subsection{Subfield distance of stabilizer codes}

A stabilizer code $Q$ of length $n$ over $\F_q$ corresponds to an additive
code $D\subseteq\F_{q^2}^n$ which is self-orthogonal with respect to the
trace-alternating form~\cite[Section~IV]{kkks}. Under our identification,
this form is given by
\[
\left\langle \varpi^{1-q}a+b,\varpi^{1-q}a'+b'\right\rangle_a
=\mathrm{tr}(b\cdot a'-b'\cdot a),
\qquad a,b,a',b'\in\F_q^n.
\]
Let $D^{\perp_a}$ denote the dual with respect to this form.
If $D$ is $\F_{q^2}$-linear, then $D^{\perp_a}=D^{\perp_h}$.
We define
\[
\dl(Q)=\min\{\wtl(\varepsilon):\varepsilon\in D^{\perp_a}\setminus D\},\qquad
d_Z(Q)=\min\{\wtH(\varepsilon):\varepsilon\in(D^{\perp_a}\setminus D)\cap\F_q^n\},
\]
and $d'_X(Q)=\min\{r(\varepsilon):\varepsilon\in D^{\perp_a}\setminus D,\ \varepsilon\notin\F_q^n\}$, $r$ being the map defined in \eqref{laR}. These are the
distances relevant for degenerate decoding. Note that, if $Q$ is a CSS code, $d_Z(Q)$ coincides with the usual $Z$-distance. If $C\subseteq\F_{q^2}^n$ is linear with $C^{\perp_h}\subseteq C$, the
quantum code has $D=C^{\perp_h}$ and $D^{\perp_a}=C$~\cite{kkks}; therefore $\dl(Q)\ge\dl(C)$ and
$d_Z(Q)\ge\dH(\subfield{C}{q})$.

A CSS code is given by linear codes $C_2\subseteq C_1\subseteq\F_q^n$, with
$D=\varpi^{1-q}C_2\oplus C_1^\perp$ and $D^{\perp_a}=\varpi^{1-q}C_1\oplus C_2^\perp$; its asymmetric distances are
$d_Z=\min\wtH(C_2^\perp\setminus C_1^\perp)$ and $d_X=\min\wtH(C_1\setminus C_2)$, and it is \emph{pure} in $Z$ if
$d_Z=\dH(C_2^\perp)$.

\begin{proposition}\label{prop:dzdx}
Let $\lambda\ge1$.
\begin{enumerate}[label=(\alph*)]
\item For every stabilizer code $Q$ as before, $\min\left\{d_Z(Q),\lambda d'_X(Q)\right\}\le\dl(Q)\le d_Z(Q)$, and
\[
\dl(Q)=\min\Big\{d_Z(Q),\ \min_{r\ge1}\big(W_r(Q)+(\lambda-1)r\big)\Big\},
\]
where $W_r(Q)=\min\{\wtH(\varepsilon):\varepsilon\in D^{\perp_a}\setminus D,\ r(\varepsilon)=r\}$.
\item For every CSS code, $\dl(Q)=\min\{d_Z,\lambda d_X\}$.
\end{enumerate}
\end{proposition}
\begin{proof}
Let us start by proving (a). If $\varepsilon\in D^{\perp_a}\setminus D$ lies in $\F_q^n$, then $\wtl(\varepsilon)=\wtH(\varepsilon)\ge d_Z(Q)$;
otherwise $\wtl(\varepsilon)\ge\lambda r(\varepsilon)\ge\lambda d'_X(Q)$. The upper bound is attained by a diagonal
element of $D^{\perp_a}\setminus D$ of weight $d_Z(Q)$, and the formula is proved as in
Proposition~\ref{prop:profile}.

Finally, let us show (b). Let $\varepsilon=\varpi^{1-q}a+b\in D^{\perp_a}\setminus D$, with $a\in C_1$, $b\in C_2^\perp$. Note that
$r(\varepsilon)=\wtH(a)$. If $a\in C_2$, then $\varpi^{1-q}a\in D$, so $b\in C_2^\perp\setminus C_1^\perp$ and
$$\wtl(\varepsilon)=\lambda\wtH(a)+\card(\operatorname{supp}b\setminus\operatorname{supp}a)\ge\wtH(b)\ge d_Z.$$ If
$a\notin C_2$, then $\wtl(\varepsilon)\ge\lambda\wtH(a)\ge\lambda d_X$. Both values are attained, by
$b\in C_2^\perp\setminus C_1^\perp$ and by $\varpi^{1-q}a$ with $a\in C_1\setminus C_2$.
\end{proof}

In a CSS code, the non-diagonal part of a logical operator can always be taken alone, so that $W_{d_X}(Q)=d_X$:
there is no coupling between the two types of errors. For a non-CSS code, the logical operators with few
non-diagonal entries may be forced to have many diagonal ones, that is, $W_r(Q)$ may be much larger than $r$, which
cannot be seen from $(d_Z,d_X)$. For instance, the $[[63,45]]_2$ code considered in Example~\ref{ex:63} below has
$\dl\ge\min\{\lambda+4,7\}$, whereas any CSS code with $d_Z=7$ and $d_X=1$ has
$\lambda$-subfield distance $\min\{\lambda,7\}$.

\subsection{When is the subfield approach better?}
We first bound the $\lambda$-subfield distance of CSS codes.

\begin{definition}\label{def:DqKq}
For $1\le k\le n$, let $D_q(n,k)$ be the largest minimum
distance of an $[n,k]_q$ linear code and $K_q(n,d)$ the largest dimension of a linear code of length $n$ and minimum
distance at least $d$.
\end{definition}

\begin{lemma}\label{lem:cssub}
Let $Q$ be a CSS code with parameters $[[n,k,d_Z/d_X]]_q$, $k\ge1$.
\begin{enumerate}[label=(\alph*)]
\item If $d_X\ge2$, then $C_1^\perp\neq\{0\}$, and therefore $\dim C_2^\perp\ge k+1$.
\item If $Q$ is pure in $Z$, then $d_Z\le D_q(n,k)$, and $d_Z\le D_q(n,k+1)$ if $d_X\ge2$.
\item If $Q$ is pure ($d_Z=\dH(C_2^\perp)$ and $d_X=\dH(C_1)$), then $d_Z\le D_q(n,k+n-K_q(n,d_X))$, and
therefore
\[
\dl(Q)\le U_q(n,k,\lambda):=\max_{1\le d\le n,\ K_q(n,d)\ge k}\ \min\big\{D_q(n,k+n-K_q(n,d)),\ \lambda d\big\},
\]
\end{enumerate}
\end{lemma}
\begin{proof}
Recall that $k=\dim C_2^\perp-\dim C_1^\perp=n-\dim C_2-\dim C_1^\perp$.

Let us prove (a). If $C_1^\perp=\{0\}$, then $C_1=\F_q^n$, and since $k\ge1$ we have $C_2\ne\F_q^n$, which implies that some unit vector $e_i$ lies in
$C_1\setminus C_2$ and $d_X=1$.

Let us see (b). The code $C_2^\perp$ is linear, of dimension $k+\dim C_1^\perp\ge k$ and minimum distance $d_Z$; then the result follows from Item (a).

Finally, we prove (c). Since the code $C_1$ has minimum distance $d_X$, then $n-\dim C_1^\perp\le K_q(n,d_X)$, and $C_2^\perp$ has dimension
$$k+\dim C_1^\perp\ge k+n-K_q(n,d_X).$$ The bound for $\dl(Q)$ follows from Proposition~\ref{prop:dzdx}(b), since the map
$D_q(n,\cdot)$ is nonincreasing.
\end{proof}

Let us state our comparative result for $\lambda$-subfield distances.

\begin{theorem}\label{thm:dominance}
Let $C\subseteq\F_{q^2}^n$ be linear with $C^{\perp_h}\subseteq C$ and $k_Q=2\dim C-n>0$, and let $Q$ be the
corresponding quantum code. Set $d=\dH(\subfield{C}{q})$ and assume that there is no $[n,k_Q+1,\ge d]_q$ linear code (i.e., $D_q(n,k_Q+1)<d$). Then,
for every CSS code $Q'$ with parameters $[[n,k_Q,d'_Z/d'_X]]_q$ which is pure in $Z$ and has $d'_Z\ge d$, and for
every $1\le\lambda<d$,
\[
\dl(Q')=\lambda<\min\{d,\ \lambda+\dH(C)-1\}\le\dl(Q),
\]
provided that $\dH(C)\ge2$.
\end{theorem}
\begin{proof}
With the notation as at the beginning of this subsection, note that if $d'_X\ge2$, Lemma~\ref{lem:cssub}(b) gives $d\le d'_Z\le D_q(n,k_Q+1)<d$, a contradiction. Hence $d'_X=1$, and
$\dl(Q')=\min\{d'_Z,\lambda\}=\lambda$ by Proposition~\ref{prop:dzdx}(b). Finally,
$\dl(Q)\ge\dl(C)\ge\min\{d,\dH(C)+\lambda-1\}$ by Theorem~\ref{thm:main}, since $\dH(\cD)\ge1$.
\end{proof}

As a consequence of the above result, it holds that if one prescribes a phase distance at least $d$, every pure CSS code of the same length and dimension is
forced to have $d_X=1$, and thus a single error involving a qudit-flip is a logical error, whereas our code still detects every single qudit-flip error at no cost in dimension or in phase distance. The hypothesis of
Theorem~\ref{thm:dominance} holds, for instance, for the $[[15,7]]_2$ code of the forthcoming Example~\ref{ex:15}, since there is
no $[15,8,5]_2$ code~\cite{codetables}. Moreover, the codes $\sfCq{\cI^+(\delta)}{q^2}$ of Section~\ref{sec:hermitian} satisfy
$\dim\sfCq{\cI^+(\delta)}{q}=k_Q$ by Proposition~\ref{prop:optimal}, and thus Theorem~\ref{thm:dominance} applies whenever there is no $[n,k_Q+1,\ge d]_q$ code, $d$ being the minimum distance of the $q$-ary BCH code of designed distance $\delta$, and $\dH(\sfCq{\cI^+(\delta)}{q^2})\ge2$. In general, Lemma~\ref{lem:cssub} and the tables of~\cite{codetables} give an upper bound for all pure CSS codes, which we will compare with our codes in
Section~\ref{sec:hermitian}. We restrict the comparison to pure CSS codes because, for them, $d_Z$ and $d_X$ are minimum distances of the linear codes $C_2^\perp$ and $C_1$, so that tables of bounds for classical linear codes give rigorous upper bounds. 
These comparisons do not cover impure CSS codes
or the non-CSS constructions of~\cite{ELS,EzermanGrassl,TBF,XZZX}.

\begin{remark}\label{rem:limits}
We will divide this remark into three parts.

(1) \emph{Pure dephasing.} For our codes and for CSS codes, the nontrivial diagonal logical Pauli operators are represented by the elements of $V\setminus V_0$, where $V_0\subseteq V$ are classical codes with $\dim V/V_0=k_Q$. When considering Hermitian codes, $V=\subfield{C}{q}$ and $V_0=\subfield{(C^{\perp_h})}{q}$, and, since $\subfield{(C^{\perp_h})}{q}=\subfield{(C^{\perp})}{q}=\Tr(C)^\perp$ by~\eqref{eq:delsarte}, Proposition~\ref{prop:trace}(b) shows that $\dim V_0=n-2\dim C+\dim\subfield{C}{q}$. A complement of $V_0$ in $V$ is an
$[n,k_Q]_q$ code all of whose nonzero words have weight at least $d_Z$, so $d_Z\le D_q(n,k_Q)$. Thus, when $p_X=0$,
neither approach can beat the best classical code.

(2) \emph{Weak bias.} For $\lambda$ close to $1$, $\dl$ is close to the minimum Hamming distance and the best symmetric codes are
preferable.

(3) \emph{Gates.} The analysis above concerns storage and transmission, that is, a
code-capacity noise model. In a fault-tolerant circuit, the bias is degraded by
non-bias-preserving gates~\cite{GM,Puri}, so syndrome extraction must use bias-preserving
gates. Such gates are already used in practice: the cat-qubit processor of~\cite{Putterman}
corrects phase flips with a repetition code whose syndromes are extracted with a
noise-biased CX gate, and the non-CSS XZZX surface code has been studied at circuit level
with Kerr-cat qubits and bias-preserving CX gates~\cite{Darmawan}. Designing
bias-preserving syndrome-extraction circuits for the codes of this paper, whose stabilizers
mix $X$-, $Y$- and $Z$-type factors, and assessing them under circuit-level biased noise
is left for future work.
\end{remark}

\section{The subfield distance of cyclic codes}\label{sec:cyclic}

We use an evaluation description of cyclic codes in terms of polynomials in one variable over an extension
field, evaluated at the roots of unity \cite{bierbrauercyclic}. In the terminology of~\cite{galindodistancestabilizer}, these are univariate $\{1\}$-affine variety codes. BCH codes appear as subfield subcodes when the complement of the defining set contains consecutive
elements.

Let $r$ be a multiple of $m$, let $n$ be a divisor of $q^r-1$ and let $\beta\in\F_{q^r}$ be
a primitive $n$-th root of unity. We denote $\Z_n=\Z/n\Z$.
Let $P=\{\beta^0,\beta^1,\dots,\beta^{n-1}\}$ be the set of
$n$-th roots of unity, ordered in this way, and let
\[
  \ev:\F_{q^r}[X]/(X^n-1)\to\F_{q^r}^n,\qquad \ev(f)=\left(f(\beta^0),\dots,f(\beta^{n-1})\right).
\]
Every class in the quotient ring $\F_{q^r}[X]/(X^n-1)$ has a unique representative
$f=\sum_{i\in\Z_n}f_iX^i$, and we call $\operatorname{supp}(f)=\{i: f_i\neq0\}\subseteq\Z_n$
its support. For every integer $k\ge0$, we set
$f^{[q^k]}:=\sum_i f_i^{q^k}X^{q^k i}$
(exponents taken modulo $n$).
If $u$ is an integer coprime to $n$ and
$S\subseteq\Z_n$, we set $uS=\{us:s\in S\}$, and we say that $S$ is \emph{$u$-closed} if
$uS=S$.

\begin{definition}\label{def:closint}
Let $u$ be coprime to $n$ and $S\subseteq\Z_n$. We denote by $\Clos{S}{u}$ the smallest
$u$-closed subset of $\Z_n$ containing $S$, and by $\Int{S}{u}$ the largest $u$-closed subset
of $\Z_n$ contained in $S$. Both are unions of $u$-cyclotomic cosets: $\Clos{S}{u}$ is the
union of those cosets \emph{meeting} $S$, and $\Int{S}{u}$ is the union of those cosets
\emph{contained in} $S$.
\end{definition}

Note that $\Int{\left(\Z_n\setminus S\right)}{u}=\Z_n\setminus\Clos{S}{u}$ and
$\Clos{\left(-S\right)}{u}=-\Clos{S}{u}$.

\begin{definition}
\label{1afine}
For $\cI\subseteq\Z_n$, we define the \textit{$\{1\}$-affine variety code}
$$\sfC{\cI}=\operatorname{Span}_{\F_{q^r}}\{\ev(X^i):i\in\cI\}=\{\ev(f):\operatorname{supp}(f)\subseteq\cI\}.$$
For $m\mid r$ we define $\sfCq{\cI}{q^m}:=\subfield{(\sfC{\cI})}{q^m}=\sfC{\cI}\cap\F_{q^m}^n$.
\end{definition}

We call $\cI$ the \emph{defining set} of the code; note that $\dim_{\F_{q^r}}\sfC{\cI}=\card{\cI}$.
When $\cI$ is $q^m$-closed, the code $\sfCq{\cI}{q^m}$ is the \textit{cyclic code} over $\F_{q^m}$ with
zeros $\{\beta^{-k}:k\notin\cI\}$ (see Remark \ref{rem:defset} below); its dimension is $\dim_{\F_{q^m}}\sfCq{\cI}{q^m}=\card{\cI}$, and its BCH-designed distance is $\Gap(\cI)+1$ (Lemma~\ref{lem:bch}).
Both codes have length $n$ (the number of evaluation points), the first one over $\F_{q^r}$ and the second one over $\F_{q^m}$. Note that in this way we can recover any cyclic code (without repeated roots) as a subfield subcode of a $\{1\}$-affine variety code. 

BCH codes are a particular family of cyclic codes: if $\Z_n\setminus\cI$ contains a run of
$\delta-1$ consecutive elements of $\Z_n$, then $\dH(\sfC{\cI})\ge\delta$ by the BCH bound (Lemma~\ref{lem:bch}), and
$\sfCq{\cI}{q}$ is a BCH code (in the usual terminology, its defining set $\Z_n\setminus(-\Int{\cI}{q})$ contains $\delta-1$ consecutive elements).

The following lemma summarizes key properties we will use later.

\begin{lemma}\label{lem:cyclic}
Let $\cI,\cJ\subseteq\Z_n$ and $m\mid r$.
\begin{enumerate}[label=(\alph*)]
\item $\ev$ is an isomorphism of $\F_{q^r}$-vector spaces; in particular $\dim_{\F_{q^r}}\sfC{\cI}=\card{\cI}$.
\item $\ev(f)^{q^k}=\ev(f^{[q^k]})$ for every $k\ge0$; hence $\sfC{\cI}^{q^k}=\sfC{q^k\cI}$.
\item $\sfC{\cI}$ and $\sfCq{\cI}{q^m}$ are cyclic codes.
\item $\sfCq{\cI}{q^m}=\sfCq{\Int{\cI}{q^m}}{q^m}$ and $\dim_{\F_{q^m}}\sfCq{\cI}{q^m}=\card{\Int{\cI}{q^m}}$.
\item If $\cI$ and $\cJ$ are $q^m$-closed, then $\sfCq{\cJ}{q^m}\subseteq\sfCq{\cI}{q^m}$ if and only if $\cJ\subseteq\cI$.
\item If $\cI$ is $q^m$-closed, then $\left(\sfCq{\cI}{q^m}\right)^{\perp}=\sfCq{\Z_n\setminus(-\cI)}{q^m}$.
\item If $\cI$ is $q^m$-closed, then $\Tr_{\F_{q^m}/\F_q}\left(\sfCq{\cI}{q^m}\right)=\sfCq{\Clos{\cI}{q}}{q}$.
\end{enumerate}
\end{lemma}

\begin{proof}
(a) The evaluation points are distinct, so a polynomial of degree $<n$ vanishing on $P$ is zero.

(b) Since $\beta^n=1$,
$f(\beta^j)^{q^k}=\sum_i f_i^{q^k}\beta^{jq^ki}=f^{[q^k]}(\beta^j)$, and $f\mapsto f^{[q^k]}$ is a bijection from the
polynomials supported on $\cI$ onto those supported on $q^k\cI$.

 (c) If $\sigma$ is the cyclic shift, then
$\sigma(\ev(f))=\ev(f(\beta^{-1}X))$, and $f(\beta^{-1}X)$ has the same support as $f$; moreover $\F_{q^m}^n$ is
$\sigma$-invariant.

(d) We begin by proving the first claim in this item. By (a) and (b), $\ev(f)\in\F_{q^m}^n$ if and only if $\ev(f)^{q^m}=\ev(f)$, if and only if $f^{[q^m]}=f$, that
is, $f_{q^mi}=f_i^{q^m}$ for all $i\in\Z_n$. In that case $\operatorname{supp}(f)$ is $q^m$-closed, so it is contained
in $\Int{\cI}{q^m}$ whenever it is contained in $\cI$. For the dimension, let
$O=\{b,q^mb,\dots,q^{m(\ell-1)}b\}$ be a $q^m$-cyclotomic coset, $\card O=\ell$. The polynomials $f$ with
$\operatorname{supp}(f)\subseteq O$ and $f^{[q^m]}=f$ are determined by $f_b$, which must satisfy
$f_b^{q^{m\ell}}=f_b$. Since $q^r\equiv1\pmod n$, we have $\ell\mid r/m$, so $\F_{q^{m\ell}}\subseteq\F_{q^r}$ and
these polynomials form an $\F_{q^m}$-space isomorphic to $\F_{q^{m\ell}}$, of dimension $\ell$. Summing over the
cosets contained in $\Int{\cI}{q^m}$ and using (a), we get $\dim_{\F_{q^m}}\sfCq{\cI}{q^m}=\card{\Int{\cI}{q^m}}$.

(e) One implication is clear. Let us see the other one by contradiction: if $j\in\cJ\setminus\cI$, its $q^m$-cyclotomic coset $O$ satisfies $O\subseteq\cJ$ and
$O\cap\cI=\emptyset$; by the proof of (d) there is $f\neq0$ with $\operatorname{supp}(f)=O$ and
$\ev(f)\in\sfCq{\cJ}{q^m}$, but $\ev(f)\notin\sfC{\cI}$ by (a).

(f) For $i,j\in\Z_n$,
$\ev(X^i)\cdot\ev(X^j)=\sum_{k=0}^{n-1}\beta^{(i+j)k}=0$ if $i+j\not\equiv0\pmod n$; hence
$\sfCq{\Z_n\setminus(-\cI)}{q^m}\subseteq(\sfCq{\cI}{q^m})^\perp$. Finally, $\cI$ and $\Z_n\setminus(-\cI)$ are
$q^m$-closed; by (d) the dimensions are $n-\card{\cI}$, which concludes the proof of this item.

Let us prove the last item, (g). Write $\Tr=\Tr_{\F_{q^m}/\F_q}$. If $c=\ev(f)\in\sfCq{\cI}{q^m}$, then by (b)
$\Tr(c)=\sum_{k=0}^{m-1}c^{q^k}=\ev\big(\sum_{k=0}^{m-1}f^{[q^k]}\big)$, whose support is contained in
$\bigcup_kq^k\cI=\Clos{\cI}{q}$ (because $\cI$ is $q^m$-closed). Hence
$\Tr(\sfCq{\cI}{q^m})\subseteq\sfCq{\Clos{\cI}{q}}{q}$. By~\eqref{eq:delsarte}, (f) and (d),
\[
\Tr(\sfCq{\cI}{q^m})^{\perp}=\sfCq{\Z_n\setminus(-\cI)}{q^m}\cap\F_q^n=\sfCq{\Int{(\Z_n\setminus(-\cI))}{q}}{q}
=\sfCq{\Z_n\setminus(-\Clos{\cI}{q})}{q},
\]
which has dimension $n-\card{\Clos{\cI}{q}}$. Thus $\Tr(\sfCq{\cI}{q^m})$ has dimension $\card{\Clos{\cI}{q}}$, which by (d) is also the dimension of $\sfCq{\Clos{\cI}{q}}{q}$ ($\Clos{\cI}{q}$ is $q$-closed); the inclusion above is therefore an equality, which concludes the proof of the lemma.
\end{proof}

\begin{remark}\label{rem:defset}
Keep the above notation. If $c=\ev(f)$ and $c(X)=\sum_jc_jX^j$, then $c(\beta^{-k})=\sum_j\sum_if_i\beta^{(i-k)j}=nf_k$. Hence, if $\cI$ is
$q^m$-closed, $\sfCq{\cI}{q^m}$ is the cyclic code over $\F_{q^m}$ whose zeros are $\{\beta^{-k}:k\notin\cI\}$, i.e.,
in the usual terminology its defining set is $\Z_n\setminus(-\cI)$. Since $\cI$ determines the code, we call $\cI$
the \emph{defining set} of $\sfCq{\cI}{q^m}$ in this paper.
\end{remark}

\begin{definition}
We define the \emph{amplitude} of a nonempty $\cI\subseteq\Z_n$ as follows:
$$\Amp(\cI):=\min\{i\in\N:\exists\, t\in\Z_n,\ \cI\subseteq\{t,t+1,\dots,t+i-1\}\},$$ and the
\emph{(biggest) gap} of $\cI$ is
$$\Gap(\cI):=\max\{i\ge 0:\exists\, t\in\Z_n,\ \{t,t+1,\dots,t+i-1\}\subseteq\Z_n\setminus \cI\}.$$
We set $\Gap(\emptyset)=n$. In particular, $\Gap(\Z_n)=0$, since $i=0$ is allowed in the definition.
\end{definition}
Since $\cI\subseteq\{t,\dots,t+i-1\}$ if and only if $\{t+i,\dots,t+n-1\}\subseteq\Z_n\setminus \cI$,
we have $\Gap(\cI)=n-\Amp(\cI)$ for $\cI\neq\emptyset$.

The following bound, known as the BCH bound, can be found in~\cite{macwilliams,huffman}:
\begin{lemma}[BCH bound]\label{lem:bch}
If $\emptyset\neq \cI\subseteq\Z_n$, then $\dH(\sfC{\cI})\ge n-\Amp(\cI)+1=\Gap(\cI)+1$.
\end{lemma}
\begin{proof}
Let $A=\Amp(\cI)$ and $\cI\subseteq\{t,\dots,t+A-1\}$. If $\operatorname{supp}(f)\subseteq\cI$,
then $f=X^tg$ with $g$ a polynomial of degree at most $A-1$ (as $\beta^{jn}=1$, exponents may
be taken modulo $n$), so $\ev(f)_j=\beta^{tj}g(\beta^j)$. If $f\neq0$, then $g\neq0$ has at most
$A-1$ roots in $P$; thus $\wtH(\ev(f))\ge n-A+1$.
\end{proof}

We now apply Theorem~\ref{thm:main} to the code $\sfCq{\cI}{q^m}\subseteq\F_{q^m}^n$ with
respect to the subfield $\F_q\subseteq\F_{q^m}$. We name $\Clos{\cI}{q}$  the $q$-{\it closure}
and $\Int{\cI}{q}$  the $q$-{\it interior} of $\cI$. We recall the notation and results introduced in Section~\ref{subsec:lowerbound} and start with a lemma.

\begin{lemma}\label{lem:Dcyclic}
Let $\cI\subseteq\Z_n$ be $q^m$-closed and $C=\sfCq{\cI}{q^m}$. Then,
\begin{enumerate}[label=(\alph*)]
\item $\cD(C)\subseteq\bcD(C)\subseteq\sfC{\cI\cup q\cI}$; hence $\dH(\cD)\ge\dH(\bcD)\ge\Gap(\cI\cup q\cI)+1$.
\item $\subfield{C}{q}=\sfCq{\Int{\cI}{q}}{q}$, which has dimension $\card{\Int{\cI}{q}}$ and $\dH(\subfield{C}{q})\ge\Gap(\Int{\cI}{q})+1$.
\end{enumerate}
\end{lemma}
\begin{proof}
(a) If $c=\ev(f)\in C$, then $c^q=\ev(f^{[q]})\in\sfC{q\cI}$ by Lemma~\ref{lem:cyclic}(b);
hence $c-c^q\in\sfC{\cI\cup q\cI}$. Now use Lemma~\ref{lem:bch} to conclude the proof.

(b) Since $\subfield{C}{q}=\sfC{\cI}\cap\F_q^n$, the statement follows by   Lemma~\ref{lem:cyclic}(d) and Lemma~\ref{lem:bch}.
\end{proof}

\begin{remark}\label{rem:Dtrace}
Note that $\cD\not\subseteq\F_q^n$ in general, so $\cD$ is not contained in the subfield subcode
$\sfCq{\Clos{\cI}{q}}{q}$ and it is contained in the $\F_{q^r}$-code $\sfC{\cI\cup q\cI}\subseteq\sfC{\Clos{\cI}{q}}$. By Lemma~\ref{lem:cyclic}(g), $\Tr(C)=\sfCq{\Clos{\cI}{q}}{q}$ and then Proposition~\ref{prop:trace} gives
$\dH(\cD)\ge\dH(\sfCq{\Clos{\cI}{q}}{q})\ge\Gap(\Clos{\cI}{q})+1$.

For $m=2$, we have
$\Clos{\cI}{q}=\cI\cup q\cI$ and $\cD=(\varpi-\varpi^q)\,\sfCq{\cI\cup q\cI}{q}$. The bound of Lemma~\ref{lem:Dcyclic}(a) is the BCH bound
of a $q$-ary cyclic code, and any better bound for this code (Hartmann--Tzeng, Roos, or its true minimum distance)
can be used instead. In this language Proposition~\ref{prop:tradeoff} says that, if $\sfCq{\cI}{q^2}$ is Hermitian
dual-containing, then $\Z_n\setminus(-\Int{\cI}{q})\subseteq\cI\cup q\cI$.
\end{remark}

Let us state the main result of this section.

\begin{theorem}\label{thm:cyclic}
Let $\lambda\ge1$ and let $\emptyset\ne \cI\subseteq\Z_n$ be $q^m$-closed. Then,
\[
  \dl\big(\sfCq{\cI}{q^m}\big)\ge\min\big\{(\lambda-1)\big(\Gap(\cI\cup q\cI)+1\big)+\Gap(\cI)+1,\ \Gap(\Int{\cI}{q})+1\big\},
\]
where the second term is omitted if $\Int{\cI}{q}=\emptyset$.
\end{theorem}
\begin{proof}
Combine Theorem~\ref{thm:main} with Lemmas~\ref{lem:bch} and~\ref{lem:Dcyclic}, using that the
bound of Theorem~\ref{thm:main} is nondecreasing in $\dH(\cD)$, $\dH(C)$ and $\dH(\subfield{C}{q})$.
\end{proof}

We conclude this section with the decoding of cyclic codes. Let $\cI\subseteq\Z_n$ be
$q^m$-closed and $C=\sfCq{\cI}{q^m}$. The BCH-type bounds of Lemmas~\ref{lem:bch}
and~\ref{lem:Dcyclic} give suitable values of $\delta_C$, $\delta_{\cD}$, and $\delta_q$ in
Proposition~\ref{prop:decoding}, and then $L$ is the lower bound of
Theorem~\ref{thm:cyclic}. The component decoders can be implemented by decoding in the
corresponding containing BCH codes and checking membership in $C$, $\cD$, and
$\subfield{C}{q}$, respectively; membership in $\cD$ and the required lift are tested by the
same linear system. For $m=2$, Proposition~\ref{prop:trace}(b) also identifies $\cD$ with a
scalar multiple of the cyclic trace code over $\F_q$. For $n=q^{r}-1$ the containing codes are
evaluation codes of polynomials with exponents in an interval of $\Z_n$ over a field of size
$n+1$ (generalized Reed--Solomon type codes), which can be decoded up to half their designed
distance in time polynomial in $n$ (for instance, by the Berlekamp--Massey algorithm); this
gives polynomial-time decoding of every error of $\lambda$-weight strictly less than half the
bound of Theorem~\ref{thm:cyclic}. Replacing a designed distance by a larger computed minimum
distance requires a component decoder reaching the corresponding larger radius.

\section{Hermitian dual-containing cyclic codes}\label{sec:hermitian}

As indicated in the introduction, the subfield metric is useful in the Hermitian setting. 
The section is organized as follows. We begin with preliminary results on Hermitian dual-containing codes and the trace code. Subsection~\ref{subsec:construction} then presents the main construction and proves the key distance lower bound. Subsection~\ref{subsec:otherlengths} extends the results to other lengths, Subsection~\ref{increasing} studies how to increase the term $\dH(\cD)$ in Theorem~\ref{thm:main} and Subsection~\ref{subsec:tables} tabulates some resulting asymmetric quantum codes.

From now on, $m=2$, $r=2s$ and $n=q^{2s}-1$; thus $q^2$-closed sets give codes over $\F_{q^2}$.
In this section, $n$ is the length of the code $\sfCq{\cI}{q^2}\subseteq\F_{q^2}^n$, of its subfield subcode
$\sfCq{\cI}{q}\subseteq\F_q^n$ and of the quantum code $[[n,2k-n]]_q$ it gives, where $k=\dim_{\F_{q^2}}\sfCq{\cI}{q^2}$;
the evaluation points live in $\F_{q^{2s}}$ and $2s$ is the order of $q$ modulo $n$. In Subsection~\ref{subsec:otherlengths},
we allow any length $n$ coprime to $q$ such that the order of $q$ modulo $n$ is even.
For $b\in\Z_n$, we denote by $\cI_b$ the $q^2$-cyclotomic coset of $b$ modulo $n$; let $\cA$
be the set of minimal elements of the $q^2$-cyclotomic cosets and $n_a=\card{\cI_a}$ for
$a\in\cA$. Note that $v\cI_b=\cI_{vb}$ for $v$ coprime to $n$, and that the $q$-cyclotomic
coset of $b$ is $\cI_b\cup\cI_{qb}$.

Hermitian dual-containing $J$-affine variety codes, and their subfield subcodes, were studied in~\cite{galindodistancestabilizer} (see also~\cite{AKS}); the following criterion is the version we need.

\begin{lemma}\label{lem:hermitian}
Let $\cI\subseteq\Z_n$ be $q^2$-closed and $C=\sfCq{\cI}{q^2}$. Then
$C^{\perp_h}=\sfCq{\Z_n\setminus(-q\cI)}{q^2}$. Consequently, the following statements are equivalent:
\begin{enumerate}[label=(\roman*)]
\item $C^{\perp_h}\subseteq C$;
\item $\Z_n=\cI\cup(-q\cI)$;
\item for every $a\in\cA$, $\cI_a\subseteq\cI$ or $\cI_{-qa}\subseteq\cI$.
\end{enumerate}
\end{lemma}
\begin{proof}
By Lemma~\ref{lem:cyclic} (f) and (b) it holds that
$C^{\perp_h}=(C^\perp)^{q}=\big(\sfCq{\Z_n\setminus(-\cI)}{q^2}\big)^q
=\sfCq{q(\Z_n\setminus(-\cI))}{q^2}=\sfCq{\Z_n\setminus(-q\cI)}{q^2}$.
By Lemma~\ref{lem:cyclic}(e), (i) is equivalent to $\Z_n\setminus(-q\cI)\subseteq\cI$, i.e.,
to (ii). Finally, $\cI_a\subseteq-q\cI$ if and only if $\cI_{-q^{-1}a}\subseteq\cI$, and
$\cI_{-q^{-1}a}=\cI_{-qa}$ because $-qa=q^2\cdot(-q^{-1}a)$; this gives the equivalence between (ii) and (iii) and the result is proved.
\end{proof}

Next we state some other useful results.

\begin{lemma}\label{lem:sym}
For $a\in\Z_n$: (i) $\cI_a=\cI_{-a}$ if and only if $\cI_{qa}=\cI_{-qa}$;
(ii) $\cI_a=\cI_{-qa}$ if and only if $\cI_{qa}=\cI_{-a}$.
\end{lemma}
\begin{proof}
Multiplication by $q$ is a bijection of $\Z_n$ mapping $\cI_b$ onto $\cI_{qb}$, and
$\cI_{-q^2a}=\cI_{-a}$. Applying it to $\cI_a=\cI_{-a}$ (respectively, $\cI_a=\cI_{-qa}$)
gives $\cI_{qa}=\cI_{-qa}$ (respectively, $\cI_{qa}=\cI_{-q^2a}=\cI_{-a}$), and multiplication
by $q^{-1}$ gives the converses.
\end{proof}

\begin{lemma}\label{lem:gcd}
Let $e\ge1$ be an odd integer and set $d=\gcd(s,e)$. Then,
$\gcd(q^{2s}-1,q^e+1)=q^d+1$ and $\gcd(q^{2s}-1,q^e-1)=q^d-1$.
\end{lemma}
\begin{proof}
Recall that $\gcd(q^a-1,q^b-1)=q^{\gcd(a,b)}-1$. Since $e$ is odd, $\gcd(2s,e)=d$, which gives
the second equality.

For the first one, let $g=\gcd(q^{2s}-1,q^e+1)$. As $e/d$ is odd,
$q^d+1\mid q^e+1$, and $q^d+1\mid q^{2d}-1\mid q^{2s}-1$; thus $q^d+1\mid g$. Conversely,
$$g\mid\gcd(q^{2s}-1,q^{2e}-1)=q^{2d}-1=(q^d-1)(q^d+1),$$ therefore $h:=g/(q^d+1)$ divides $q^d-1$
and divides $(q^e+1)/(q^d+1)$. Writing $y=q^d$ and $k=e/d$ (odd),
$(y^k+1)/(y+1)=\sum_{i=0}^{k-1}(-y)^i\equiv1\pmod{y-1}$, so $h=1$.
\end{proof}

Let $\tau$ be the {\it largest odd divisor of $s$}, and consider the following two sets:
\[
  \Gamma:=\{b\in\Z_n: \cI_b=\cI_{-qb}\} \mbox{ and } \Phi:=\{b\in\Z_n: \cI_b=\cI_{qb}\}.
\]

\begin{proposition}\label{prop:gamma}
Keep the notation introduced in this section. The set  $\Gamma$ is the subgroup of $\Z_n$ generated by $\gamma:=\dfrac{q^{2s}-1}{q^\tau+1}$ and the set
$\Phi$ is the subgroup generated by $\dfrac{q^{2s}-1}{q^\tau-1}$. In particular, the following hold:
\begin{enumerate}[label=(\alph*)]
\item $\gamma$ is the smallest nonzero element of $\Gamma$, and $\gamma\in\cA$;
\item if $s$ is odd, then $\gamma=q^s-1$;
\item every nonzero $b\in\Phi$ satisfies $b\ge\frac{n}{q^\tau-1}>\gamma$.
\end{enumerate}
\end{proposition}
\begin{proof}
As the order of $q$ modulo $n$ is $2s$, $b\in\Gamma$ if and only if $-qb\equiv q^{2k}b$ for
some $k$, which holds if and only if $(q^e+1)b\equiv0\pmod n$ for
some odd $e\in\{1,3,\dots,2s-1\}$. For a fixed $e$, the solutions form the subgroup generated
by $n/\gcd(n,q^e+1)=n/(q^{d_e}+1)$, with $d_e=\gcd(s,e)$, by Lemma~\ref{lem:gcd}. Since
$d_e$ is an odd divisor of $s$, $d_e\mid\tau$ and $\tau/d_e$ is odd, so $q^{d_e}+1\mid
q^\tau+1$ and the subgroup for $e$ is contained in the subgroup generated by $n/(q^\tau+1)$,
which is attained for $e=\tau$. Hence $\Gamma=\langle\gamma\rangle$. The same argument, with
$qb\equiv q^{2k}b$, $q^e-1$ instead of $q^e+1$ and $q^{d_e}-1\mid q^\tau-1$, gives the
claim for $\Phi$.

Recall that, for any positive divisor $t$ of $n$, the representatives
of $\langle t\rangle$ in $\{0,\dots,n-1\}$ are $0,t,\dots,n-t$.
Thus, if $t<n$, its smallest nonzero representative is $t$. This proves (c), and we are ready to prove Items (a) and (b).

(a) Clearly $\gamma$ is the smallest nonzero element of $\langle\gamma\rangle$; a subgroup of
$\Z_n$ is closed under multiplication by integers, so $\cI_\gamma\subseteq\Gamma$, and then
$\min\cI_\gamma\ge\gamma$, i.e., $\gamma\in\cA$. To prove (b), note that since $s$ is odd, $\tau=s$ and
$\gamma=(q^{2s}-1)/(q^s+1)=q^s-1$.
\end{proof}

\begin{remark}
For $s$ even, $\tau=\max\{j\text{ odd}: q^j+1\mid q^s-1\}$, since for odd $j$ we have
$q^j+1\mid q^s-1$ if and only if $j\mid s$. The parity of $\tau$ is essential: for instance,
for $q=2$ and $s=2$ we have $n/(q^s+1)=3$, but $\cI_3=\{3,12\}\neq\cI_{-6}=\{6,9\}$, while
$\gamma=n/(q+1)=5$.
\end{remark}

\begin{corollary}\label{cor:limit}
If $\Delta\subseteq\Z_n$ is $q^2$-closed and $\left(\sfCq{\Delta}{q^2}\right)^{\perp_h}\subseteq\sfCq{\Delta}{q^2}$,
then $\Gamma\subseteq\Int{\Delta}{q}$. Consequently, $\Gap(\Int{\Delta}{q})\le\gamma-1$.
\end{corollary}
\begin{proof}
Let $b\in\Gamma$. Since $\cI_b=\cI_{-qb}$, Lemma~\ref{lem:hermitian}(iii) forces
$\cI_b\subseteq\Delta$; as $qb\in\Gamma$, also $\cI_{qb}\subseteq\Delta$. Thus the
$q$-cyclotomic coset $\cI_b\cup\cI_{qb}$ of $b$ is contained in $\Delta$, i.e.,
$b\in\Int{\Delta}{q}$. Since $\Gamma=\{0,\gamma,2\gamma,\dots\}$ and $\gamma\mid n$, every
interval of $\gamma$ consecutive elements of $\Z_n$ meets $\Gamma$.
\end{proof}

Hence, the BCH-type bound of Lemma~\ref{lem:Dcyclic}(b) can guarantee
$\dH(\subfield{C}{q})\ge\delta$ for a Hermitian dual-containing code
only when $\delta\le\gamma$. We now construct such codes for every
$\delta\le\gamma$.

\subsection{\texorpdfstring{A construction with $\dH(\subfield{C}{q})\ge\delta$}{A construction with large subfield-subcode distance}}\label{subsec:construction}
Keep the above notation. For $a\in\Z_n$ let
\[
  T_a:=\{\cI_a,\cI_{qa},\cI_{-qa},\cI_{-a}\},\qquad U_a:=\cI_a\cup\cI_{qa}\cup\cI_{-qa}\cup\cI_{-a}.
\]
$U_a$ is the orbit of $a$ under the group generated by multiplication by $q$ and by $-1$, so
the sets $U_a$ form a partition of $\Z_n$. Let $\cA':=\{a\in\Z_n: a=\min U_a\}\subseteq\cA$;
then $\Z_n=\bigsqcup_{a\in\cA'}U_a$. The group $\{1,q,-q,-1\}$ modulo $q^2$ acts on the
$q^2$-cyclotomic cosets and $T_a$ is the orbit of $\cI_a$, so $\card{T_a}\in\{1,2,4\}$. Let us study the different possibilities regarding the cardinality of the sets $T_a$.

\begin{lemma}\label{lem:cases}
For $a\in\cA'$, exactly one of the following holds:
\begin{enumerate}[label=(\arabic*)]
\item $\card{T_a}=1$;
\item $\card{T_a}=2$ and $\cI_a=\cI_{-qa}$ (then $\cI_{qa}=\cI_{-a}$);
\item $\card{T_a}=2$ and $\cI_a=\cI_{-a}$ (then $\cI_{qa}=\cI_{-qa}$);
\item $\card{T_a}=2$ and $\cI_a=\cI_{qa}$ (then $\cI_{-a}=\cI_{-qa}$);
\item $\card{T_a}=4$.
\end{enumerate}
Moreover, Case (1) occurs exactly for $a=0$ and, if $q$ is odd, for $a=n/2$; in Case (2) we
have $a\ge\gamma$; and in Case (4) we have $a>\gamma$.
\end{lemma}
\begin{proof}
If $\card{T_a}=2$, the stabilizer of $\cI_a$ in $\{1,q,-q,-1\}$ has index $2$ and therefore contains
exactly one of the values $q,-q,-1$ (if it contained two of them it would contain the third and
$\card{T_a}=1$). The statements in brackets follow from Lemma~\ref{lem:sym} (or directly from
the stabilizer being a subgroup). Case (1) means $\cI_a=\cI_{qa}=\cI_{-qa}$ (then also
$\cI_{-a}=q\cI_{-qa}=q\cI_a=\cI_a$), i.e., $a\in\Gamma\cap\Phi$. By
Proposition~\ref{prop:gamma},
$\Gamma\cap\Phi=\langle n/(q^\tau+1)\rangle\cap\langle n/(q^\tau-1)\rangle=\langle n/g\rangle$
with $g=\gcd(q^\tau+1,q^\tau-1)$, which is $1$ if $q$ is even and $2$ if $q$ is odd. In Case
(2), $0\ne a\in\Gamma$, and in Case (4), $0\neq a\in\Phi$; now use
Proposition~\ref{prop:gamma} to finish the proof.
\end{proof}

Fix $1\le\delta\le\gamma$ and set $[\delta-1]:=\{1,\dots,\delta-1\}$. We want to obtain a $q^2$-closed
set $\cI(\delta)$ such that $\sfCq{\cI(\delta)}{q^2}$ is Hermitian dual-containing
(Lemma~\ref{lem:hermitian}) and $\Int{\cI(\delta)}{q}\cap[\delta-1]=\emptyset$, which gives
$\Gap(\Int{\cI(\delta)}{q})\ge\delta-1$ and $\dH(\sfCq{\cI(\delta)}{q})\ge\delta$. The conditions of
Lemma~\ref{lem:hermitian}(iii) only relate $q^2$-cosets inside the same $U_a$, and the
$q$-cyclotomic cosets are contained in the sets $U_a$; therefore we can choose $\cI(\delta)\cap U_a$
independently for each $a\in\cA'$. This is done in the next Algorithm~\ref{alg:delta}; the variant
$\cI^+(\delta)$ includes every coset that can be added without violating the constraints.

Theorem~\ref{thm:construction} is the main result of this subsection. It shows that Algorithm~\ref{alg:delta} determines sets $\cI(\delta)$ and $\cI^+(\delta)$, which give rise to sets $\Delta$ for which a lower bound for $\dl(\sfCq{\Delta}{q^2})$ is easy to express.
\algnewcommand{\Output}{\item[\textbf{Output:}]}
\begin{algorithm}[ht]
\caption{Construction of $\cI(\delta)$ and $\cI^+(\delta)$}\label{alg:delta}
\begin{algorithmic}[1]
\Require $q$, $s$, $n=q^{2s}-1$, $1\le\delta\le\gamma$
\Output $\cI(\delta)$ and $\cI^+(\delta)$
\State $\cI\gets\emptyset$, $\cI^+\gets\emptyset$
\For{$a\in\cA'$}
  \If{Case (1) or (2)} \State $\cI\gets\cI\cup\cI_a\cup\cI_{qa}$; \quad $\cI^+\gets\cI^+\cup\cI_a\cup\cI_{qa}$
  \ElsIf{Case (3)} \State $\cI\gets\cI\cup\cI_a$; \quad $\cI^+\gets\cI^+\cup\cI_a$, and also $\cup\,\cI_{qa}$ if $a\ge\delta$
  \ElsIf{Case (4)} \State $\cI\gets\cI\cup\cI_a$; \quad $\cI^+\gets\cI^+\cup\cI_a\cup\cI_{-a}$
  \ElsIf{$(\cI_{-qa}\cup\cI_{-a})\cap[\delta-1]\neq\emptyset$} \Comment{Case (5a)}
     \State $\cI\gets\cI\cup\cI_a\cup\cI_{-a}$; \quad $\cI^+\gets\cI^+\cup\cI_a\cup\cI_{-a}$
  \Else \Comment{Case (5b)}
     \State $\cI\gets\cI\cup\cI_a\cup\cI_{-a}\cup\cI_{-qa}$; \quad $\cI^+\gets\cI^+\cup\cI_a\cup\cI_{-a}\cup\cI_{-qa}$, and also $\cup\,\cI_{qa}$ if $a\ge\delta$
  \EndIf
\EndFor
\State \Return $\cI(\delta)=\cI$, $\cI^+(\delta)=\cI^+$
\end{algorithmic}
\end{algorithm}

To state this theorem, let $S_1,\dots,S_4\subseteq\cA'$ be the sets of $a\in\cA'$ in Cases (1)--(4) of Lemma~\ref{lem:cases}, respectively, and let
$S_5$ (respectively $S_6$) be the set of $a\in\cA'$ with $\card{T_a}=4$ such that
$(\cI_{-qa}\cup\cI_{-a})\cap[\delta-1]\neq\emptyset$ (respectively $=\emptyset$). Then
$\cA'=\bigsqcup_{i=1}^6S_i$. By Lemma~\ref{lem:cases}, $S_1=\{0\}$ if $q$ is even and
$S_1=\{0,n/2\}$ if $q$ is odd. Note also that every $a\in S_5$ satisfies $a<\delta$, since
$a=\min U_a$.

\begin{theorem}\label{thm:construction}
Let $1\le\delta\le\gamma$ and let $\Delta\in\{\cI(\delta),\cI^+(\delta)\}$ be obtained from Algorithm~\ref{alg:delta}. Then:
\begin{enumerate}[label=(\roman*)]
\item $\Delta$ is $q^2$-closed and $\left(\sfCq{\Delta}{q^2}\right)^{\perp_h}\subseteq\sfCq{\Delta}{q^2}$;
\item $\Int{\Delta}{q}\cap[\delta-1]=\emptyset$; hence $\Gap(\Int{\Delta}{q})\ge\delta-1$ and $\dH(\sfCq{\Delta}{q})\ge\delta$;
\item $\dim_{\F_{q^2}}\sfCq{\cI(\delta)}{q^2}=\sum_{a\in S_1}n_a+2\sum_{a\in S_2}n_a+\sum_{a\in S_3\cup S_4}n_a+2\sum_{a\in S_5}n_a+3\sum_{a\in S_6}n_a$, and
$\dim_{\F_{q^2}}\sfCq{\cI^+(\delta)}{q^2}=\dim_{\F_{q^2}}\sfCq{\cI(\delta)}{q^2}+\sum_{a\in S_4}n_a+\sum_{a\in S_3\cup S_6,\ a\ge\delta}n_a$;
\item for every $\lambda\ge1$,
\[
  \dl(\sfCq{\Delta}{q^2})\ge\min\big\{(\lambda-1)\big(\Gap(\Delta\cup q\Delta)+1\big)+\Gap(\Delta)+1,\ \delta\big\}\ge\min\{\lambda+\Gap(\Delta),\delta\}.
\]
\end{enumerate}
\end{theorem}
\begin{proof}
(i) $\Delta$ is a union of $q^2$-cosets. By Lemma~\ref{lem:hermitian}(iii) we must check, for
each $a\in\cA'$ and each $c\in U_a$, that $\cI_c\subseteq\Delta$ or $\cI_{-qc}\subseteq\Delta$.
The pairs $\{\cI_c,\cI_{-qc}\}$ with $c\in U_a$ are $\{\cI_a,\cI_{-qa}\}$ and
$\{\cI_{qa},\cI_{-a}\}$ (as $\cI_{-q^2a}=\cI_{-a}$ and $\cI_{q^2a}=\cI_a$). In Case (1) and
in Case (2), everything in $U_a$ is included. In Case (3), $\cI_{-qa}=\cI_{qa}$, so both pairs
equal $\{\cI_a,\cI_{qa}\}$ and $\cI_a$ is included. In Case (4), both pairs equal
$\{\cI_a,\cI_{-a}\}$ and $\cI_a$ is included. In Case (5), $\cI_a$ and $\cI_{-a}$ are
included, one from each pair. This proves the criterion for $\cI(\delta)$, and thus, for $\cI^+(\delta)$ too.

(ii) We check that no $q$-cyclotomic coset meeting $[\delta-1]$ is contained in $\Delta$. In
Case (1), $U_a=\{a\}$ with $a\in\{0,n/2\}$; in Case (2), $a\ge\gamma\ge\delta$ by
Lemma~\ref{lem:cases}; in both cases $U_a\cap[\delta-1]=\emptyset$ because $a=\min U_a$ (and
$n/2>\gamma\ge\delta$). In Case (4), $a>\gamma\ge\delta$ by Lemma~\ref{lem:cases}, and again
$U_a\cap[\delta-1]=\emptyset$. In Case (3), $U_a=\cI_a\cup\cI_{qa}$ is a single $q$-coset;
it is contained in $\Delta$ only for $\cI^+$ and $a\ge\delta$, in which case
$U_a\cap[\delta-1]=\emptyset$. In Case (5) the $q$-cosets in $U_a$ are
$Q_1=\cI_a\cup\cI_{qa}$ and $Q_2=\cI_{-a}\cup\cI_{-qa}$. $Q_1\subseteq\Delta$ only for
$\cI^+$ and $a\ge\delta$ (then $U_a\cap[\delta-1]=\emptyset$), and $Q_2\subseteq\Delta$ only
in Case (5b), where $Q_2\cap[\delta-1]=\emptyset$. The bound
$\dH(\sfCq{\Delta}{q})\ge\delta$ follows from Lemma~\ref{lem:Dcyclic}(b).

(iii) Follows from the construction and Lemma~\ref{lem:cyclic}(d), since all $q^2$-cosets in
$U_a$ have $n_a$ elements.

(iv) Theorem~\ref{thm:cyclic} together with (ii) shows this item; the last inequality uses
$\Gap(\Delta\cup q\Delta)\ge0$.
\end{proof}

\begin{remark}
For $n=q^{2s}-1$, adjoining evaluation at $0$ gives the
parity extension $\widehat C$ of
$C=\sfCq{\Delta}{q^2}$ (sometimes called an extended cyclic code, see \cite{campsCSST}).
Extension commutes with taking subfield subcodes and preserves
Hermitian dual-containment for the codes constructed above. Indeed, $0\in\Delta$ implies $\mathbf{1}\in C$, so every
$z\in C^{\perp_h}$ has coordinate sum zero. Since $n=-1$
in $\F_{q^2}$, the parity extension satisfies
\[
\widehat C^{\perp_h}
=(C^{\perp_h}\times\{0\})
\oplus\langle(\mathbf{1},1)\rangle_{\F_{q^2}}
\subseteq\widehat C.
\]
Moreover, $\dH(\subfield{\widehat C}{q})\ge\delta+1$ by the
extended BCH bound, and $\dH(\cD(\widehat C))\ge2$, since
these words have coordinate sum zero. Hence
\[
  \dl(\widehat C)\ge
  \min\{2\lambda+\Gap(\Delta)-1,\delta+1\}.
\]
If $\dim_{\F_{q^2}}C=k$, the resulting quantum code has
parameters $[[n+1,2k-n-1]]_q$. 

% Proofs: write $\widehat C=\{(\ev(f),f_0):\operatorname{supp}(f)\subseteq\Delta\}$ (the univariate $\emptyset$-affine variety code with exponent set $\Delta$, i.e., evaluation at all of $\F_{q^{2s}}$). Since $\sum_j\ev(f)_j=nf_0=-f_0$ (as $n\equiv-1\pmod p$), $\widehat C=\{(c,-\sum_jc_j):c\in C\}$ is the parity extension; hence every word of $\widehat C$ has coordinate sum zero, so every nonzero word of $\cD(\widehat C)$ has weight at least $2$, and the extension of a word of $\F_q^n$ lies in $\F_q^{n+1}$, which gives the commutation with subfield subcodes. Moreover, $\widehat C^{\perp_h}=\{(y,y_\infty):y-y_\infty\mathbf{1}\in C^{\perp_h}\}$, which is contained in $\widehat C$ if and only if $\mathbf{1}\in C$ and $\sum_jz_j=0$ for all $z\in C^{\perp_h}=\sfCq{\Z_n\setminus(-q\Delta)}{q^2}$; both conditions are equivalent to $0\in\Delta$, which holds because $\Z_n=\Delta\cup(-q\Delta)$ (Lemma~\ref{lem:hermitian}). Finally, if $0\neq c=\ev(f)\in\subfield{C}{q}$, then $\operatorname{supp}(f)\subseteq\Int{\Delta}{q}\subseteq\{0,\delta,\dots,n-1\}$, so $g(x)=f_0+\sum_{i\ge\delta}f_ix^{n-i}$ is a nonzero polynomial of degree at most $n-\delta$ whose values at the points of $\F_{q^{2s}}$ ($x=0$ and $x=\beta^{-j}$) are the coordinates of $(c,f_0)$ in some order; as $g$ has at most $n-\delta$ roots, $\wtH(c,f_0)\ge q^{2s}-(n-\delta)=\delta+1$. Together with $\dH(\widehat C)\ge\dH(C)\ge\Gap(\Delta)+1$ and Theorem~\ref{thm:main}, this gives the bound.
\end{remark}

We supplement Theorem~\ref{thm:construction} with the following proposition, which shows that our choice of $\cI^+(\delta)$ is optimal.

\begin{proposition}\label{prop:optimal}
Keep the above notation. Let $1\le\delta\le\gamma$. Among all $q^2$-closed sets $\Delta\subseteq\Z_n$ such that
$\sfCq{\Delta}{q^2}$ is Hermitian dual-containing and $\Int{\Delta}{q}\cap[\delta-1]=\emptyset$,
the set $\cI^+(\delta)$ has the largest cardinality. Moreover
$\Int{\cI^+(\delta)}{q}=\Int{\Z_n\setminus[\delta-1]}{q}$ and
\[
  2\dim_{\F_{q^2}}\sfCq{\cI^+(\delta)}{q^2}-n=\dim_{\F_q}\sfCq{\cI^+(\delta)}{q}=\card{\Int{\Z_n\setminus[\delta-1]}{q}},
\]
that is, the quantum code obtained from $\sfCq{\cI^+(\delta)}{q^2}$ has as many logical qudits
as the $q$-ary BCH code $\sfCq{\Int{\Z_n\setminus[\delta-1]}{q}}{q}$ of designed distance
$\delta$, and this is the maximum possible.
\end{proposition}
\begin{proof}
Let $Q_1$ and $Q_2$ be the sets introduced in the proof of Theorem~\ref{thm:construction}. As explained above the constraints split over the orbits $U_a$, $a\in\cA'$, and thus we bound
$\card{\Delta\cap U_a}$ for each $a$. In Cases (1) and (2) all of $U_a$ is forced by
Lemma~\ref{lem:hermitian}(iii) (in Case (2), use that $\cI_a=\cI_{-qa}$ and
$\cI_{qa}=\cI_{-q\cdot qa}$). In Case (3), if $a<\delta$ the $q$-coset $U_a$ meets $[\delta-1]$,
so at most one of $\cI_a,\cI_{qa}$ can be taken. In Case (5), if $a<\delta$ then
$Q_1\not\subseteq\Delta$; therefore at most three cosets are taken, and three cosets force
$Q_2\subseteq\Delta$, which is impossible if $Q_2$ meets $[\delta-1]$, i.e., if $a\in S_5$.
In all remaining situations, $\cI^+(\delta)$ contains all of $U_a$. Comparing with
Algorithm~\ref{alg:delta}, $\cI^+(\delta)\cap U_a$ attains the maximum for every $a$.

The same case analysis shows that $\cI^+(\delta)$ contains every $q$-coset disjoint from
$[\delta-1]$, then $$\Int{\cI^+(\delta)}{q}=\Int{\Z_n\setminus[\delta-1]}{q}.$$ Moreover, in each
orbit, $2\card{\cI^+(\delta)\cap U_a}-\card{U_a}=\card{\Int{\cI^+(\delta)}{q}\cap U_a}$:
this is $n_a$ in Case (1); $2n_a$ in Cases (2), (4), and (3) with $a\ge\delta$; zero in Case
(3) with $a<\delta$ and in Case (5a); $2n_a$ in Case (5b) with $a<\delta$; and $4n_a$ in
Case (5) with $a\ge\delta$. Summing over $a\in\cA'$ gives the displayed equality. Finally, for
any admissible $\Delta$, Lemma~\ref{lem:dimCq} gives
$2\card{\Delta}-n\le\dim_{\F_q}\sfCq{\Delta}{q}=\card{\Int{\Delta}{q}}\le\card{\Int{\Z_n\setminus[\delta-1]}{q}}$.
\end{proof}

\begin{remark}
With the above notation, recall that $\cI(\delta)\subseteq\cI^+(\delta)$, and then $\Gap(\cI^+(\delta))\le\Gap(\cI(\delta))$ and
$$\Gap(\cI^+(\delta)\cup q\cI^+(\delta))\le\Gap(\cI(\delta)\cup q\cI(\delta));$$ the larger
dimension of $\cI^+(\delta)$ may come at the cost of a weaker first term within $\min$ in
Theorem~\ref{thm:construction}(iv). Table~\ref{tab:delta} shows that, for $\delta=\gamma$, the
gain in dimension can be large while $g$ and $\Gap(\Int{\Delta}{q})$ do not change (only $g'$ may drop).
\end{remark}

Table~\ref{tab:delta} below compares $\Delta=\cI(\gamma)$ and $\Delta^+=\cI^+(\gamma)$ for small parameters. Here
$k=\card{\Delta}$ is the dimension of the code over $\F_{q^2}$, $k_Q=2k-n$ is the quantum
dimension, $g=\Gap(\Delta)$ and $g'=\Gap(\Delta\cup q\Delta)$ (and similarly $k^+$, $k_Q^+$, $g^+$, $g'^+$ for $\Delta^+$). In all cases
$\Gap(\Int{\Delta}{q})=\Gap(\Int{\Delta^+}{q})=\gamma-1$.

\begin{table}[ht]
\centering
\caption{Comparison of $\cI(\gamma)$ and $\cI^+(\gamma)$.\label{tab:delta}}
\begin{tabular}{rrrr|rrrr|rrrr}
\toprule
$q$ & $s$ & $n$ & $\gamma$ & $k$ & $k_Q$ & $g$ & $g'$ & $k^+$ & $k_Q^+$ & $g^+$ & $g'^+$\\
\midrule
2&2&15&5 & 11&7&2&0 & 11&7&2&0\\
2&3&63&7 & 48&33&3&1 & 54&45&3&0\\
2&4&255&85 & 151&47&14&0 & 151&47&14&0\\
2&5&1023&31 & 768&513&15&1 & 948&873&15&0\\
3&2&80&20 & 57&34&7&0 & 57&34&7&0\\
3&3&728&26 & 547&366&9&1 & 677&626&9&0\\
4&2&255&51 & 182&109&13&1 & 187&119&13&0\\
5&2&624&104 & 451&278&21&1 & 473&322&21&0\\
\bottomrule
\end{tabular}
\end{table}

For the codes of Theorem~\ref{thm:construction}, $\delta_{\cD}\le2$ in all the cases of
Table~\ref{tab:delta}, so in Proposition~\ref{prop:decoding} every correctable error with a
non-diagonal entry is already decoded in step~(i), and steps~(ii) and~(iii) only add the
decoding of diagonal errors.

\begin{example}\label{ex:15}
Let $q=2$ and $s=2$, so $n=15$ and $\gamma=5$. Take $\delta=5$ and $C=\sfCq{\cI(5)}{4}\subseteq\F_4^{15}$ (here
$\cI^+(5)=\cI(5)$). Then $\Gap(\cI(5))=2$, $\Gap(\cI(5)\cup2\cI(5))=0$, and $C$ is a
$[15,11,3]_4$ code. Enumerating all its codewords, the minimal BR-weights (with respect to the
componentwise order) are
\[
  (5,0),\ (2,1),\ (1,2),\ (0,3),
\]
so $\dl(C)=\min\{5,2+\lambda,1+2\lambda,3\lambda\}=\min\{5,2+\lambda\}$ for $\lambda\ge1$, and
$\dH(\cD)=1$, $\dH(\subfield{C}{2})=5$. Theorem~\ref{thm:construction} gives
$\dl(C)\ge\min\{5,2+\lambda\}$, so the bound is sharp for every $\lambda\ge1$.

The $4$-cyclotomic cosets modulo $15$ are $\{0\}$, $\{1,4\}$, $\{2,8\}$,
$\{3,12\}$, $\{5\}$, $\{6,9\}$, $\{7,13\}$, $\{10\}$, $\{11,14\}$, and Algorithm~\ref{alg:delta} gives
$\cI(5)=\{0,1,3,4,5,7,10,11,12,13,14\}$, which misses only $\{2,8\}$ and $\{6,9\}$. Hence $k=\card{\cI(5)}=11$
and the quantum code has $k_Q=2k-n=7$: it is a $[[15,7]]_2$ code. The $2$-cyclotomic cosets modulo $15$ are
$\{0\}$, $\{1,2,4,8\}$, $\{3,6,9,12\}$, $\{5,10\}$, $\{7,11,13,14\}$, and the ones contained in $\cI(5)$ are
$\{0\}$, $\{5,10\}$ and $\{7,11,13,14\}$, so $\Int{\cI(5)}{2}=\{0,5,7,10,11,13,14\}$ has $7$ elements and
$\Gap(\Int{\cI(5)}{2})=4$: $\subfield{C}{2}$ is the binary $[15,7,5]$ BCH code. Since there is no
$[15,8,5]_2$ code~\cite{codetables}, Theorem~\ref{thm:dominance} applies: every pure CSS $[[15,7]]_2$ code with $d_Z\ge5$ has
$d_X=1$ and $\lambda$-subfield distance $\lambda$, whereas the $[[15,7]]_2$ code obtained from $C$ has
$\lambda$-subfield distance at least $\min\{5,\lambda+2\}$. Without any restriction on $d_Z$,
Lemma~\ref{lem:cssub} and~\cite{codetables} give, for every pure CSS $[[15,7]]_2$ code,
$\dl\le U_2(15,7,\lambda)=\max\{\min\{5,\lambda\},\min\{4,2\lambda\},\min\{3,3\lambda\},2\}$, that is,
$\dl\le3$ for $\lambda\le3/2$, $\dl\le2\lambda$ for $3/2\le\lambda\le2$, $\dl\le4$ for $2\le\lambda\le4$ and $\dl\le\lambda$
for $4\le\lambda\le5$. Hence our code is better for $1<\lambda<2$ and $2<\lambda<5$ (for $\lambda=2$ the upper bound equals our value).

\end{example}

\begin{example}\label{ex:63}
Let $q=2$ and $s=3$, so $n=q^{2s}-1=63$ (the evaluation points are the $63$rd roots of unity of $\F_{64}$) and
$\gamma=7$. Take $\delta=7$ and $C=\sfCq{\cI^+(7)}{4}\subseteq\F_4^{63}$, a $[63,54]_4$ code (Table~\ref{tab:delta}),
which gives a $[[63,45]]_2$ quantum code. Its subfield subcode $\subfield{C}{2}$ is the $[63,45,7]$ binary BCH code, and
$\Gap(\cI^+(7)\cup 2\cI^+(7))=0$, so $\Tr(C)=\F_2^{63}$ and $\dH(\cD)=1$. The designed distance of $C$ is
$\Gap(\cI^+(7))+1=4$, but a computation with Magma gives $\dH(C)=5$, and Inequality~\eqref{eq:maintrace} proves
\[
  \dl(C)\ge\min\{\lambda+4,7\}.
\]
This bound is attained for every $\lambda\ge1$. $C$ has codewords of weight $5$ with BR-weight $(4,1)$, hence with
$\lambda$-weight $\lambda+4$, and $\subfield{C}{2}$ has codewords of weight $7$ (this agrees with the values $11/2$, $6$ and $7$ for $\lambda=3/2$, $2$ and $3$ obtained with Proposition~\ref{prop:image}). In addition,
Lemma~\ref{lem:cssub} and the tables of~\cite{codetables} show that every pure CSS $[[63,45]]_2$ code has
$\lambda$-subfield distance at most $9/2$, $5$ and $6$, respectively.

The values of $\dl$ are compared in the following table.
\begingroup
\setlength{\topsep}{10pt}
\begin{center}
\begin{tabular}{@{}lccc@{}}
\toprule
$\lambda$ & $3/2$ & $2$ & $3$\\
\midrule
Distance $d_\lambda$ of our $[[63,45]]_2$ code & $11/2$ & $6$ & $7$\\
Upper bound on $\dl$ for pure CSS $[[63,45]]_2$ codes & $9/2$ & $5$ & $6$\\
\bottomrule
\end{tabular}
\end{center}
\endgroup
Concretely, consider a channel in which $99\%$ of the errors are phase flips and the total error probability is
between $5\%$ and $10\%$; then $\lambda$ is between $2.79$ and $3.40$ (Remark~\ref{rem:lambdavalues}), and
$\dl(C)=\min\{\lambda+4,7\}>6$ in this range. Hence our code corrects every error of $\lambda$-weight less than $\min\{\lambda+4,7\}/2$, which exceeds both $3$ and $\lambda$ for $2<\lambda<7/2$;
that is, every pattern of up to three phase flips and, separately, every single error involving a bit flip, i.e., an $X$ or $Y$ error on a single qubit. A pure CSS
$[[63,45]]_2$ code correcting the same errors would need $d_Z\ge7$ and $d_X\ge3$; then $C_1$ would be a code of
minimum distance at least $3$, so $\dim C_1^\perp\ge6$, and $C_2^\perp$ would be a $[63,\ge51,\ge7]_2$ code, which does not
exist (the largest minimum distance of a $[63,51]_2$ code is $5$~\cite{codetables}).
\end{example}

\subsection{Other lengths}\label{subsec:otherlengths}
The construction presented earlier in this section does not depend on $n$ being exactly $q^{2s}-1$. Let us state the values of $n$ for which this construction remains valid. Let $n$ be coprime to $q$ and assume that
the order of $q$ modulo $n$ is even, say $2s$; then $n\mid q^{2s}-1$, the $n$-th roots of unity lie in $\F_{q^{2s}}$, the codes
$\sfCq{\cI}{q^2}$ have length $n$ over $\F_{q^2}$, and all the notions of the previous subsections ($\cI_b$, $\cA'$,
$T_a$, $U_a$, the cases (1)--(5) of Lemma~\ref{lem:cases} (but not its last sentence, which uses $n=q^{2s}-1$), $\Gamma$, $\Phi$) make sense. Let $\tau$ be the largest odd divisor
of $s$.

\begin{proposition}\label{prop:gammageneral}
$\Gamma$ is the subgroup of $\Z_n$ generated by $n/\gcd(n,q^\tau+1)$, and $\Phi$ the subgroup generated
by $n/\gcd(n,q^\tau-1)$.
\end{proposition}
\begin{proof}
As in the proof of Proposition~\ref{prop:gamma}, $b\in\Gamma$ if and only if $(q^e+1)b\equiv0\pmod n$
for some odd $e\in\{1,\dots,2s-1\}$, and the solutions for a fixed $e$ form the subgroup generated by
$n/\gcd(n,q^e+1)$. Since $n\mid q^{2s}-1$, Lemma~\ref{lem:gcd} gives
$\gcd(n,q^e+1)=\gcd(n,q^{d_e}+1)$ with $d_e=\gcd(s,e)$, which divides $\gcd(n,q^\tau+1)$ because
$q^{d_e}+1\mid q^\tau+1$. The same argument works for $\Phi$.
\end{proof}

Let $\gamma(n)$ be the smallest nonzero element of $\Gamma$ (with $\gamma(n)=n$ if $\Gamma=\{0\}$), define
\[
  \phi^*(n)=\min\{\max\{\min\cI_a,\min\cI_{-a}\}:\ 0\neq a\in\Phi,\ \cI_a\neq\cI_{-a}\}
  \qquad(\min\emptyset=n),
\]
and $\theta(n)=\min\{\gamma(n),\phi^*(n)\}$. For $n=q^{2s}-1$ we have $\phi^*(n)>\gamma(n)$ by
Proposition~\ref{prop:gamma}(c), and then $\theta(n)=\gamma(n)$. In general $\phi^*(n)$ may be smaller than
$\gamma(n)$ (for instance $\gamma(51)=51$ and $\phi^*(51)=34$ for $q=4$). Table~\ref{tab:lengths} lists these
quantities for several lengths.

\begin{table}[ht]
\centering
\caption{Lengths $n$ with even order $2s$ of $q$ modulo $n$, and the bounds $\gamma(n)$, $\phi^*(n)$ and $\theta(n)$.
}\label{tab:lengths}
\begin{tabular}{@{}cccccccc@{}}
\toprule
$q$ & $n$ & $q^{2s}-1$ & $2s$ & $\tau$ & $\gamma(n)$ & $\phi^*(n)$ & $\theta(n)$\\
\midrule
$2$ & $15$  & $15$  & $4$ & $1$ & $5$  & $15$  & $5$\\
$2$ & $51$  & $255$ & $8$ & $1$ & $17$ & $51$  & $17$\\
$2$ & $63$  & $63$  & $6$ & $3$ & $7$  & $27$  & $7$\\
$2$ & $255$ & $255$ & $8$ & $1$ & $85$ & $255$ & $85$\\
$3$ & $20$  & $80$  & $4$ & $1$ & $5$  & $20$  & $5$\\
$3$ & $80$  & $80$  & $4$ & $1$ & $20$ & $80$  & $20$\\
$4$ & $15$  & $15$  & $2$ & $1$ & $3$  & $10$  & $3$\\
$4$ & $51$  & $255$ & $4$ & $1$ & $51$ & $34$  & $34$\\
$4$ & $85$  & $255$ & $4$ & $1$ & $17$ & $85$  & $17$\\
\bottomrule
\end{tabular}
\end{table}

The following result explains how Algorithm~\ref{alg:delta} and Theorem~\ref{thm:construction} can be adapted to the situation in this subsection.

\begin{theorem}\label{thm:generaln}
Let $1\le\delta\le n-1$. There is a $q^2$-closed set $\Delta\subseteq\Z_n$ such that $\sfCq{\Delta}{q^2}$ is
Hermitian dual-containing and $\Int{\Delta}{q}\cap[\delta-1]=\emptyset$ if and only if $\delta\le\theta(n)$.
In that case such sets are produced by Algorithm~\ref{alg:delta} (with input $n$ and $1\le\delta\le\theta(n)$) with the following change in line 8 corresponding to Case (4) of Lemma~\ref{lem:cases}: if $a<\delta$, the coset $\cI_{-a}$ is added instead of $\cI_a$ (to both $\cI(\delta)$ and $\cI^+(\delta)$).
The resulting $\cI^+(\delta)$ has the largest cardinality among all such sets, and
Theorem~\ref{thm:construction}(i), (ii) and (iv) hold.
\end{theorem}
\begin{proof}
We begin by proving the necessity of the condition $\delta\le\theta(n)$. Let $\Delta$ be such a set. If $0\ne b\in\Gamma$, then $\cI_b=\cI_{-qb}$ and $\cI_{qb}=\cI_{-q\cdot qb}$
force $\cI_b\cup\cI_{qb}\subseteq\Delta$ by Lemma~\ref{lem:hermitian}(iii); therefore the $q$-cyclotomic coset of
$b$ is contained in $\Int{\Delta}{q}$ and $b\ge\delta$; hence $\delta\le\gamma(n)$. If $0\ne a\in\Phi$ and
$\cI_a\ne\cI_{-a}$, then $\cI_a$ and $\cI_{-a}=\cI_{-qa}$ are $q$-closed, and Lemma~\ref{lem:hermitian}(iii)
requires one of them to belong to $\Delta$, hence to $\Int{\Delta}{q}$. As a consequence, one of them is disjoint from
$[\delta-1]$, i.e., $\delta\le\max\{\min\cI_a,\min\cI_{-a}\}$. Thus $\delta\le\phi^*(n)$.

Let us now see the sufficiency of that condition. The proof of Theorem~\ref{thm:construction} applies verbatim, except for the bounds on
$a$ given by Lemma~\ref{lem:cases}, which we replace as follows. In Case (1) with $a\ne0$, and in
Case (2), $a\in\Gamma$, so $a\ge\gamma(n)\ge\delta$. In Case (4), $a\in\Phi$ and $\cI_a\ne\cI_{-a}$ (otherwise
$\card{T_a}=1$); if $a\ge\delta$ nothing changes, and if $a<\delta$, then $\min\cI_a=a<\delta\le\phi^*(n)\le\max\{\min\cI_a,\min\cI_{-a}\}$, hence $\min\cI_{-a}\ge\phi^*(n)\ge\delta$, the
only $q$-closed subset of $U_a\cap\Delta$ is $\cI_{-a}$, which misses $[\delta-1]$, and both pairs
$\{\cI_a,\cI_{-qa}\}$, $\{\cI_{qa},\cI_{-a}\}$ equal $\{\cI_a,\cI_{-a}\}$; therefore the Hermitian condition holds. The
maximality of $\card{\cI^+(\delta)}$ follows orbitwise as in the proof of Proposition~\ref{prop:optimal}, where, in Case (4) with $a<\delta$, $\cI_a$ can never be taken.
\end{proof}

\begin{remark}
In Case (4) with $a<\delta$, we have $\card{\cI^+(\delta)\cap U_a}=n_a$ while $\cI_{-a}\subseteq\Int{\cI^+(\delta)}{q}$.
Thus, the equality $2\dim\sfCq{\cI^+(\delta)}{q^2}-n=\dim\sfCq{\cI^+(\delta)}{q}$ of
Proposition~\ref{prop:optimal} may fail for such lengths.
\end{remark}

We conclude this subsection by presenting two examples related to our previous results.

\begin{example}\label{ex:51}
(a) Let $q=2$ and $n=51$. The order of $2$ modulo $51$ is $8$, so $s=4$, $\tau=1$ and $51\mid2^8-1=255$; therefore the
evaluation points are $51$st roots of unity of $\F_{256}$; here $\gamma(51)=17=\theta(51)$. For $\delta=9$ the set
$\cI^+(9)$ has $39$ elements, $C=\sfCq{\cI^+(9)}{4}$ is a $[51,39]_4$ code and it gives
a $[[51,27]]_2$ quantum code. The code $\subfield{C}{2}$ has parameters $[51,27,9]_2$, $\Gap(\cI^+(9)\cup2\cI^+(9))=0$ (thus, $\dH(\cD)=1$),
and Magma gives $\dH(C)=6$. Therefore, $$\dl(C)\ge\min\{\lambda+5,9\},$$ that is, $13/2$ for $\lambda=3/2$ and $9$
for $\lambda=4$, whereas Lemma~\ref{lem:cssub} and~\cite{codetables} bound the $\lambda$-subfield distance of
every pure CSS $[[51,27]]_2$ code by $6$ and $8$, respectively.

(b) Let $q=4$ and $n=51$. The order of $4$ modulo $51$ is $4$, $s=2$ and $51\mid4^4-1=255$ (the codes have
length $51$ over $\F_{16}$). Now, $\Phi=\{0,17,34\}$ and $\cI_{17}\ne\cI_{34}$. Then, $\phi^*(51)=34<\gamma(51)=51$ and,
by Theorem~\ref{thm:generaln}, the largest $\delta$ is $34$, not $51$. For $\delta=34$, our construction gives
$k=\card{\cI^+(34)}=28$ and a $[[51,5]]_4$ quantum code. The set $\Int{\cI^+(34)}{4}$ has $6$ elements, and $\subfield{C}{4}$ is a
$[51,6,34]_4$ code. Therefore,  $\dim\subfield{C}{4}=6>k_Q$ and the equality of Proposition~\ref{prop:optimal} fails, as
anticipated in the previous remark. Moreover $\Gap(\cI^+(34)\cup4\cI^+(34))=1$ gives $\dH(\cD)\ge2$, and a computation with Magma shows that $\dH(\Tr(C))=2$, so
$\dH(\cD)=2$.
\end{example}

\subsection{\texorpdfstring{Increasing $\dH(\cD)$}{Increasing d(D)}}
\label{increasing}
In Theorem~\ref{thm:construction}(iv), $\Gap(\Delta\cup q\Delta)$ is usually $0$ or $1$. To improve the first term, we can remove suitable cosets so that $\Delta \cup q\Delta$ misses an integer interval $[t] = \{1, \dots, t\}$, and subsequently restore the dual-containing property. This subsection presents an algorithm for this construction and establishes a lower bound on the $\lambda$-subfield distance of the corresponding subfield subcode.

Let us begin by stating the above-mentioned algorithm, Algorithm~\ref{alg:enlarge}.

\begin{algorithm}[ht]
\caption{Clearing $[t]$ in $\Delta\cup q\Delta$}\label{alg:enlarge}
\begin{algorithmic}[1]
\Require A $q^2$-closed set $\Delta$ with $\sfCq{\Delta}{q^2}$ Hermitian dual-containing; $t\ge1$
\Output $\widetilde\Delta$ or \textsc{Fail}
\State $R\gets\Clos{[t]\cup q^{-1}[t]}{q^2}$
\If{there is $c\in R$ with $\cI_{-qc}\subseteq R$} \Return \textsc{Fail} \EndIf
\State \Return $\widetilde\Delta:=(\Delta\setminus R)\cup(-qR)$
\end{algorithmic}
\end{algorithm}

\begin{proposition}\label{prop:enlarge}
Let $\Delta$ and $t$ be as in Algorithm~\ref{alg:enlarge}.
\begin{enumerate}[label=(\alph*)]
\item If $\Delta'$ is a $q^2$-closed set, $\sfCq{\Delta'}{q^2}$ is Hermitian dual-containing and
$(\Delta'\cup q\Delta')\cap[t]=\emptyset$, then $\Delta'\cap R=\emptyset$ and $-qR\subseteq\Delta'$.
In particular, if the algorithm fails, no such $\Delta'$ exists.
\item If the algorithm does not fail, $\widetilde\Delta$ is $q^2$-closed, $\sfCq{\widetilde\Delta}{q^2}$ is
Hermitian dual-containing, $(\widetilde\Delta\cup q\widetilde\Delta)\cap[t]=\emptyset$, and for every
$\lambda\ge1$
\[
  \dl\big(\sfCq{\widetilde\Delta}{q^2}\big)\ge\min\big\{(\lambda-1)(t+1)+\Gap(\widetilde\Delta)+1,\ \Gap(\Int{\widetilde\Delta}{q})+1\big\}.
\]
\end{enumerate}
\end{proposition}
\begin{proof}
Let us prove Item (a). The set $(\Delta'\cup q\Delta')\cap[t]$ is empty if and only if
$\Delta'\cap([t]\cup q^{-1}[t])=\emptyset$, if and only if $\Delta'\cap R=\emptyset$, since
$\Delta'$ is $q^2$-closed. Then, for each $c\in R$, Lemma~\ref{lem:hermitian}(iii) shows that
$\cI_{-qc}\subseteq\Delta'$. Note that if $\cI_{-qc}\subseteq R$ for some $c\in R$, this contradicts
$\Delta'\cap R=\emptyset$.

We finish by showing Item (b). The set $\widetilde\Delta$ is a union of $q^2$-cosets, and $\widetilde\Delta\cap R=\emptyset$ because the
algorithm did not fail; hence $(\widetilde\Delta\cup q\widetilde\Delta)\cap[t]=\emptyset$ as in (a).
Let $c\in\Z_n$. If $c\in R$, then $\cI_{-qc}\subseteq\widetilde\Delta$. If $c\notin R$ and
$\cI_c\subseteq\Delta$, one has $\cI_c\subseteq\widetilde\Delta$. If $c\notin R$ and
$\cI_c\not\subseteq\Delta$, it holds $\cI_{-qc}\subseteq\Delta$. As a consequence, either $\cI_{-qc}\cap
R=\emptyset$ and $\cI_{-qc}\subseteq\widetilde\Delta$, or $\cI_{-qc}\subseteq R$ and then
$\cI_c=\cI_{-q\cdot(-qc)}\subseteq-qR\subseteq\widetilde\Delta$. By
Lemma~\ref{lem:hermitian}, $\sfCq{\widetilde\Delta}{q^2}$ is Hermitian dual-containing. Finally,
$\Gap(\widetilde\Delta\cup q\widetilde\Delta)\ge t$, and the bound follows from
Theorem~\ref{thm:cyclic}.
\end{proof}

Note that $\Int{\widetilde\Delta}{q}$ may be larger than $\Int{\Delta}{q}$ (the cosets in $-qR$ may
complete a $q$-coset); therefore $\Gap(\Int{\widetilde\Delta}{q})$ has to be computed. For $q=2$, $t=2$, $s\ge2$ and $\Delta\in\{\cI(\delta),\cI^+(\delta)\}$ ($\delta\ge2$ in the second case) the algorithm does not fail, since $-2R=\cI_{-1}\cup\cI_{-2}\subseteq\Delta$ (see the proof of Corollary~\ref{cor:q2}); it may fail for other sets $\Delta$.

\begin{corollary}\label{cor:q2}
Let $q=2$, $s\ge2$, $1\le\delta\le\gamma$, and let $\Delta=\cI(\delta)$ (or
$\Delta=\cI^+(\delta)$ with $\delta\ge2$). Set $\Delta':=\Delta\setminus\cI_1$. Then,
$\sfCq{\Delta'}{4}$ is Hermitian dual-containing,
$\dim_{\F_4}\sfCq{\Delta'}{4}=\dim_{\F_4}\sfCq{\Delta}{4}-s$, $\Int{\Delta'}{2}=\Int{\Delta}{2}$,
and for every $\lambda\ge1$
\[
  \dl(\sfCq{\Delta'}{4})\ge\min\{3(\lambda-1)+\Gap(\Delta')+1,\ \delta\}.
\]
\end{corollary}
\begin{proof}
The elements of $\cI_1\cup\cI_2$ are the powers $2^j$, $0\le j<2s$, which are pairwise
distinct modulo $n$ and have binary weight $1$, while the elements of $\cI_{-1}\cup\cI_{-2}$
are the values $n-2^j$, of binary weight $2s-1\ge3$. Hence, with the notation introduced before Lemma~\ref{lem:cases}, $\card{T_1}=4$, $1\in\cA'$, and
$$\min(\cI_{-1}\cup\cI_{-2})=n-2^{2s-1}=2^{2s-1}-1\ge(4^s-1)/3\ge\gamma\ge\delta.$$ Thus
$1\in S_6$ and $\cI_1,\cI_{-1},\cI_{-2}\subseteq\Delta$, $\cI_2\cap\Delta=\emptyset$ (for
$\cI^+(\delta)$ this uses $1<\delta$). We apply Algorithm~\ref{alg:enlarge} with $t=2$: since
$2^{-1}=2^{2s-1}\in\cI_2$, we get $R=\cI_1\cup\cI_2$ and $-2R=\cI_{-2}\cup\cI_{-1}\subseteq\Delta$,
therefore the algorithm does not fail and $\widetilde\Delta=\Delta\setminus\cI_1=\Delta'$. The order of
$4$ modulo $4^s-1$ is $s$ and then  $\card{\cI_1}=s$. Since $\cI_2\not\subseteq\Delta$, the
$2$-coset $\cI_1\cup\cI_2$ was not contained in $\Delta$, so $\Int{\Delta'}{2}=\Int{\Delta}{2}$
and $\Gap(\Int{\Delta'}{2})\ge\delta-1$ by Theorem~\ref{thm:construction}. Finally, applying
Proposition~\ref{prop:enlarge} we conclude the proof.
\end{proof}

\begin{example}\label{ex:enlarge}
We continue with Example~\ref{ex:63}: $q=2$, $s=3$, $n=63$, $\delta=7$ and $\Delta=\cI^+(7)$, so that $C=\sfCq{\Delta}{4}$
is a $[63,54]_4$ code with $\dl(C)=\min\{\lambda+4,7\}$. By Corollary~\ref{cor:q2}, $\Delta'=\Delta\setminus\cI_1$
(here $\card{\cI_1}=3$) gives a Hermitian dual-containing code $C'=\sfCq{\Delta'}{4}$ with parameters $[63,51]_4$, hence a $[[63,39]]_2$ quantum code. We have $\Int{\Delta'}{2}=\Int{\Delta}{2}$, hence $\subfield{C'}{2}$ is again the
binary $[63,45,7]$ BCH code, and $\Gap(\Delta')=3$ and $\Gap(\Delta'\cup2\Delta')=2$. Thus, $\dH(\cD)\ge3$ and
Corollary~\ref{cor:q2} proves
\[
  \dl(C')\ge\min\{3\lambda+1,7\}.
\]
Here the true minimum distance is larger than the designed one: Magma gives $\dH(C')=6$ and $\dH(\Tr(C'))=3$, so
that~\eqref{eq:maintrace} improves the bound to $\dl(C')\ge\min\{3\lambda+3,7\}$, and, since $\dH(\subfield{C'}{2})=7$,
$\dl(C')=7$ for every $\lambda\ge4/3$ (consistently with the values $\dl(C')=7$ for $\lambda=3/2$, $2$ and $3$ obtained with Proposition~\ref{prop:image}).

We have lost $6$ logical qubits ($45\to39$), and in exchange the
distance $7$ is reached already at $\lambda=4/3$ instead of $\lambda=3$. For $4/3\le\lambda<7/4$ every pure CSS
$[[63,39]]_2$ code has $\lambda$-subfield distance at most $4\lambda<7$ (Lemma~\ref{lem:cssub} and~\cite{codetables}).
Therefore, our code is better in this range, which contains values of $\lambda$ given by the realistic biases of
Table~\ref{tab:lambda}. Finally, note that for larger $\lambda$ the CSS bound reaches $7$.
\end{example}

Our last subsection is devoted to presenting two tables with the parameters and $\lambda$-subfield distance of codes obtained through our procedure, along with a comparison with those obtained via the CSS method.

\subsection{\texorpdfstring{Tables of parameters}{Tables of parameters}}\label{subsec:tables}

Table~\ref{tab:asym} lists asymmetric parameters of quantum codes obtained from
$C=\sfCq{\cI^+(\gamma)}{q^2}$ (of length $n=q^{2s}-1$ over $\F_{q^2}$). In all these cases $\Gap(\cI^+(\gamma)\cup q\cI^+(\gamma))=0$ and
Theorem~\ref{thm:construction}(iv) gives $\dl(C)\ge\min\{\lambda+g,\gamma\}$, where
$g=\Gap(\cI^+(\gamma))$. Hence, $\dl(C)\ge\gamma$ for $\lambda\ge\lambda_0:=\gamma-g$. By
the error-correction interpretation recalled in Section~\ref{sec:prelim} (correction of a set $E_1$ and detection of a set $E_2$), which also applies to the quantum code since $\dl(Q)\ge\dl(C)$, the quantum code corrects up to
$t_Z=\lfloor(\gamma-1)/2\rfloor$ phase-shift errors in the absence
of qudit-flip errors. Alternatively, it corrects up to
$t'_Z=\lfloor(g-1)/2\rfloor$ phase-shift errors, while detecting
any pattern consisting of up to $t'_Z$ phase-shift errors together
with one qudit-flip error.
As a symmetric code, it corrects $\lfloor g/2\rfloor$ arbitrary errors, since
$\dH(C)\ge g+1$. In Table~\ref{tab:asym} we have replaced $g+1$ by a better lower bound $d'$ for $\dH(C)$ (the true minimum
distance computed with Magma for $n=63$, the Brouwer--Zimmermann lower bound $\dH(C)\ge9$ for $n=80$ (Magma only certifies $9\le\dH(C)\le11$), and the BCH bound of the equivalent code with defining set
$2\cI^+(\gamma)$ for $q=4$), so that $\lambda_0=\gamma-d'+1$, $t'_Z=\lfloor(d'-2)/2\rfloor$ and
$\lfloor(d'-1)/2\rfloor$ arbitrary errors are corrected. For $[[80,34]]_3$ the formula gives $\lambda_0=12$; the entry $5^{*}$ in Table~\ref{tab:asym} is obtained from Proposition~\ref{prop:profile}, since
$W_1(C)\ge16$ and $W_2(C)\ge12$ give $\dl(C)\ge20$ already for $\lambda\ge5$. As explained in
Remark~\ref{rem:lambdavalues}, the values of $\lambda_0$ in the table correspond, except for the first rows, to very
strong bias.

\begin{table}[ht]
\centering
\small
\caption{Asymmetric quantum codes.}\label{tab:asym}
\begin{tabular}{@{}cccccccc@{}}
\toprule
$q$ & code & $\dH(C)\ge$ & arbitrary errors & $\lambda_0$ & $\dl(C)\ge$ & $t_Z$ & $t'_Z$\\
\midrule
$2$ & $[[15,7]]_2$ & $3$ & $1$ & $3$ & $5$ & $2$ & $0$\\
$2$ & $[[63,45]]_2$ & $5$ & $2$ & $3$ & $7$ & $3$ & $1$\\
$2$ & $[[255,47]]_2$ & $15$ & $7$ & $71$ & $85$ & $42$ & $6$\\
$2$ & $[[1023,873]]_2$ & $16$ & $7$ & $16$ & $31$ & $15$ & $7$\\
$3$ & $[[80,34]]_3$ & $9$ & $4$ & $5^{*}$ & $20$ & $9$ & $3$\\
$3$ & $[[728,626]]_3$ & $10$ & $4$ & $17$ & $26$ & $12$ & $4$\\
$4$ & $[[255,119]]_4$ & $26$ & $12$ & $26$ & $51$ & $25$ & $12$\\
$5$ & $[[624,322]]_5$ & $22$ & $10$ & $83$ & $104$ & $51$ & $10$\\
\bottomrule
\end{tabular}
\end{table}

\begin{table}[ht]
\centering
\small
\caption{Codes constructed using Theorem \ref{thm:generaln} with larger
$\lambda$-subfield distance than every pure CSS code of the same
length and dimension.}\label{tab:wins}
\begin{tabular}{@{}cccccl@{}}
\toprule
$q$ & $n$ & $\delta$ & code & $\dH(C)$, $\dH(\subfield{C}{q})$ & $(\lambda;\ \text{ours},\ \text{CSS})$\\
\midrule
$2$ & $15$  & $2$  & $[[15,11]]_2$  & $2$, $3$  & $(2;\ 3,\ 2)$, $(5/2;\ 3,\ 5/2)$\\
$2$ & $15$  & $5$  & $[[15,7]]_2$   & $3$, $5$  & $(3/2;\ 7/2,\ 3)$, $(3;\ 5,\ 4)$\\
$2$ & $51$  & $9$  & $[[51,27]]_2$  & $6$, $9$  & $(3/2;\ 13/2,\ 6)$, $(4;\ 9,\ 8)$\\
$2$ & $51$  & $11$ & $[[51,19]]_2$  & $9$, $14$ & $(1;\ 9,\ 8)$, $(5/2;\ 21/2,\ 10)$, $(6;\ 14,\ 12)$\\
$2$ & $63$  & $5$  & $[[63,51]]_2$  & $3$, $5$  & $(3;\ 5,\ 4)$\\
$2$ & $63$  & $7$  & $[[63,45]]_2$  & $5$, $7$  & $(1;\ 5,\ 4)$, $(2;\ 6,\ 5)$, $(3;\ 7,\ 6)$\\
$2$ & $255$ & $5$  & $[[255,239]]_2$ & $3$, $5$ & $(3;\ 5,\ 4)$\\
$2$ & $255$ & $7$  & $[[255,231]]_2$ & $4$, $7$ & $(5/2;\ 11/2,\ 5)$\\
$3$ & $20$  & $4$  & $[[20,12]]_3$  & $3$, $4$  & $(3/2;\ 7/2,\ 3)$\\
$3$ & $80$  & $4$  & $[[80,72]]_3$  & $3$, $4$  & $(1;\ 3,\ 2)$, $(3/2;\ 7/2,\ 3)$\\
$4$ & $15$  & $3$  & $[[15,11]]_4$  & $3$, $3$  & $(1;\ 3,\ 2)$\\
$4$ & $51$  & $6$  & $[[51,35]]_4$  & $6$, $9$  & $(3/2;\ 13/2,\ 6)$, $(4;\ 9,\ 8)$\\
$4$ & $85$  & $6$  & $[[85,69]]_4$  & $5$, $7$  & $(5/4;\ 21/4,\ 5)$\\
\bottomrule
\end{tabular}
\end{table}

Table~\ref{tab:wins} lists selected codes obtained from
Theorem~\ref{thm:generaln} by considering $q\le4$, all $n\le255$ for
which $q$ has even order modulo $n$, and all admissible $\delta$.
Each triple in the last column gives $\lambda$, our lower bound on
$\dl$, and the upper bound $U_q(n,k_Q,\lambda)$ from
Lemma~\ref{lem:cssub}(c) for pure CSS codes of the same length and
dimension. In each triple, our lower bound exceeds the CSS upper bound.
The gains are small in absolute
terms, but
they occur for the realistic values of $\lambda$ of
Remark~\ref{rem:lambdavalues}; for larger $\delta$ our bounds are dominated by the weak first term, and the CSS upper
bounds of Lemma~\ref{lem:cssub} are not tight. For larger lengths, the picture is different and our
bounds do not show an advantage at realistic biases.

\section*{Statements and declarations}

\subsection*{Use of generative AI}
During the preparation of this work the authors used Claude (Anthropic), a large language
model, to draft some proofs, to write the computer
programs with which some statements and examples
were verified and to improve the language. All the results, proofs, computations and text were
subsequently reviewed and edited by the authors, who take full responsibility for the content
of the publication.

\subsection*{Funding}
This work was partially funded by Spanish MICIU/AEI/10.13039/501100011033 and by ERDF/EU (grants PID2022-138906NB-C21 and PID2022-138906NB-C22). The third author was also partially supported by the Department of Education, Junta de Castilla y Le\'on and FEDER Funds under Grant CLU-2025-1-02-IMUVA. The fourth author was also partially supported by the NSF grant DMS-2401558, the Commonwealth Cyber Initiative, and an AMS-Simons Travel Grant.

\subsection*{Competing interests}
The authors have no competing interests to declare.

% \bibliographystyle{abbrv}
% \bibliography{BIBR}

\begin{thebibliography}{10}

\bibitem{AKS}
S.~A. Aly, A.~Klappenecker, and P.~K. Sarvepalli.
\newblock On quantum and classical {BCH} codes.
\newblock {\em IEEE Trans. Inform. Theory}, 53(3):1183--1188, 2007.

\bibitem{relativehull}
S.~E. Anderson, E.~Camps-Moreno, H.~H. López, G.~L. Matthews, D.~Ruano, and I.~Soprunov.
\newblock Relative hulls and quantum codes.
\newblock {\em IEEE Trans. Inform. Theory}, 70(5):3190--3201, 2024.

\bibitem{Aru}
F.~Arute et~al.
\newblock Quantum supremacy using a programmable superconducting processor.
\newblock {\em Nature}, 574(7779):505--510, 2019.

\bibitem{AshKni01}
A.~Ashikhmin and E.~Knill.
\newblock Nonbinary quantum stabilizer codes.
\newblock {\em IEEE Trans. Inform. Theory}, 47(7):3065--3072, 2001.

\bibitem{BCW}
J.~Bariffi, G.~Cavicchioni, and V.~Weger.
\newblock Coarsest {F}ourier-reflexive partitions for the {L}ee, homogeneous and subfield metric.
\newblock {\em ArXiv 2409.11926}, 2024.

\bibitem{bierbrauercyclic}
J.~Bierbrauer.
\newblock The theory of cyclic codes and a generalization to additive codes.
\newblock {\em Des. Codes Cryptogr.}, 25(2):189--206, 2002.

\bibitem{bierbrauerquantum}
J.~Bierbrauer and Y.~Edel.
\newblock Quantum twisted codes.
\newblock {\em J. Combin. Des.}, 8(3):174--188, 2000.

\bibitem{XZZX}
J.~P. Bonilla~Ataides, D.~K. Tuckett, S.~D. Bartlett, S.~T. Flammia, and B.~J. Brown.
\newblock The {XZZX} surface code.
\newblock {\em Nat. Commun.}, 12(1):Paper No. 2172, 2021.

\bibitem{magma}
W.~Bosma, J.~Cannon, and C.~Playoust.
\newblock The {M}agma algebra system. {I}. {T}he user language.
\newblock {\em J. Symbolic Comput.}, 24(3-4):235--265, 1997.
\newblock Computational algebra and number theory (London, 1993).

\bibitem{Calder2}
A.~R. Calderbank, E.~M. Rains, P.~W. Shor, and N.~J.~A. Sloane.
\newblock Quantum error correction and orthogonal geometry.
\newblock {\em Phys. Rev. Lett.}, 78(3):405--408, 1997.

\bibitem{css1}
A.~R. Calderbank, E.~M. Rains, P.~W. Shor, and N.~J.~A. Sloane.
\newblock Quantum error correction via codes over {${\rm GF}(4)$}.
\newblock {\em IEEE Trans. Inform. Theory}, 44(4):1369--1387, 1998.

\bibitem{calderbankgoodquantum}
A.~R. Calderbank and P.~W. Shor.
\newblock Good quantum error-correcting codes exist.
\newblock {\em Phys. Rev. A}, 54(2):1098--1105, 1996.

\bibitem{campsCSST}
E.~Camps-Moreno, H.~H. L\'opez, G.~L. Matthews, D.~Ruano, R.~San-Jos\'e, and I.~Soprunov.
\newblock An algebraic characterization of binary {CSS-T} codes and cyclic {CSS-T} codes for quantum fault tolerance.
\newblock {\em Quantum Inf. Process.}, 23(6):Paper No. 230, 2024.

\bibitem{Camps-Moreno2025-xo}
E.~Camps-Moreno, H.~H. L{\'o}pez, G.~L. Matthews, and R.~San-Jos{\'e}.
\newblock The weight hierarchy of decreasing norm-trace codes.
\newblock {\em Des. Codes Cryptogr.}, 93(7):2873--2894, July 2025.

\bibitem{christensennested}
R.~B. Christensen and O.~Geil.
\newblock On nested code pairs from the {H}ermitian curve.
\newblock {\em Finite Fields Appl.}, 68:101742, 2020.

\bibitem{Darmawan}
A.~S. Darmawan, B.~J. Brown, A.~L. Grimsmo, D.~K. Tuckett, and S.~Puri.
\newblock Practical quantum error correction with the {XZZX} code and {K}err-cat qubits.
\newblock {\em PRX Quantum}, 2(3):Paper No. 030345, 2021.

\bibitem{delsarte}
P.~Delsarte.
\newblock On subfield subcodes of modified {R}eed-{S}olomon codes.
\newblock {\em IEEE Trans. Inform. Theory}, 21(5):575--576, 1975.

\bibitem{EJLP}
M.~F. Ezerman, S.~Jitman, S.~Ling, and D.~V. Pasechnik.
\newblock {CSS}-like constructions of asymmetric quantum codes.
\newblock {\em IEEE Trans. Inform. Theory}, 59(10):6732--6754, 2013.

\bibitem{EzermanGrassl}
M.~F. Ezerman and M.~Grassl.
\newblock Asymmetric quantum codes detecting a single amplitude error.
\newblock In {\em Proc. IEEE Int. Symp. Inform. Theory (ISIT)}, 2013.
\newblock DOI 10.1109/ISIT.2013.6620361.

\bibitem{ELS}
M.~F. Ezerman, S.~Ling, and P.~Sol{\'e}.
\newblock Additive asymmetric quantum codes.
\newblock {\em IEEE Trans. Inform. Theory}, 57(8):5536--5550, 2011.

\bibitem{FLX}
K.~Feng, S.~Ling, and C.~Xing.
\newblock Asymptotic bounds on quantum codes from algebraic geometry codes.
\newblock {\em IEEE Trans. Inform. Theory}, 52(3):986--991, 2006.

\bibitem{galindodistancestabilizer}
C.~Galindo, O.~Geil, F.~Hernando, and D.~Ruano.
\newblock On the distance of stabilizer quantum codes from {$J$}-affine variety codes.
\newblock {\em Quantum Inf. Process.}, 16(4):Paper No. 111, 2017.

\bibitem{galindoimprovednestedcodepairs}
C.~Galindo, O.~Geil, F.~Hernando, and D.~Ruano.
\newblock Improved constructions of nested code pairs.
\newblock {\em IEEE Trans. Inform. Theory}, 64(4):2444--2459, 2018.

\bibitem{gahe}
C.~Galindo and F.~Hernando.
\newblock On the generalization of the construction of quantum codes from {H}ermitian self-orthogonal codes.
\newblock {\em Des. Codes Cryptogr.}, 90(5):1103--1112, 2022.

\bibitem{galindostabilizer}
C.~Galindo, F.~Hernando, and D.~Ruano.
\newblock Stabilizer quantum codes from {$J$}-affine variety codes and a new {S}teane-like enlargement.
\newblock {\em Quantum Inf. Process.}, 14(9):3211--3231, 2015.

\bibitem{Gottesman}
D.~Gottesman.
\newblock Class of quantum error-correcting codes saturating the quantum {H}amming bound.
\newblock {\em Phys. Rev. A}, 54(3):1862--1868, 1996.

\bibitem{Got97}
D.~Gottesman.
\newblock {\em Stabilizer codes and quantum error correction}.
\newblock PhD thesis, California Institute of Technology, 1997.
\newblock ArXiv quant-ph/9705052.

\bibitem{Grassl06}
M.~Grassl.
\newblock Searching for linear codes with large minimum distance.
\newblock In W.~Bosma and J.~Cannon, editors, {\em Discovering mathematics with {M}agma}, volume~19 of {\em Algorithms Comput. Math.}, pages 287--313. Springer, Berlin, 2006.

\bibitem{codetables}
M.~Grassl.
\newblock {Bounds on the minimum distance of linear codes and quantum codes}.
\newblock Available online at \url{https://www.codetables.de}, 2007.
\newblock Accessed on 2026-10-06.

\bibitem{GBR04}
M.~Grassl, T.~Beth, and M.~R\"{o}tteler.
\newblock On optimal quantum codes.
\newblock {\em Int. J. Quantum Inf.}, 2(1):55--64, 2004.

\bibitem{GHW}
M.~Grassl, A.-L. Horlemann, and V.~Weger.
\newblock The subfield metric and its application to quantum error correction.
\newblock {\em J. Algebra Appl.}, 23(7):Paper No. 2550063, 2024. ArXiv 2212.00431.

\bibitem{GM}
J.~Guillaud and M.~Mirrahimi.
\newblock Repetition cat qubits for fault-tolerant quantum computation.
\newblock {\em Phys. Rev. X}, 9(4):Paper No. 041053, 2019.

\bibitem{GuoEtAlLee}
J.~Guo, Y.~Han, A.~Chakraborty, S.~P. Jain, T.~Liu, V.~V. Albert, and A.~Lucas.
\newblock Quantum codes in the {Lee} metric.
\newblock {\em ArXiv 2610.04834}, 2026.

\bibitem{huffman}
W.~C. Huffman and V.~Pless.
\newblock {\em Fundamentals of error-correcting codes}.
\newblock Cambridge University Press, Cambridge, 2003.

\bibitem{ioffe}
L.~Ioffe and M.~M\'ezard.
\newblock Asymmetric quantum error-correcting codes.
\newblock {\em Phys. Rev. A}, 75(3):Paper No. 032345, 2007.

\bibitem{XingC}
L.~Jin, S.~Ling, J.~Luo, and C.~Xing.
\newblock Application of classical {H}ermitian self-orthogonal {MDS} codes to quantum {MDS} codes.
\newblock {\em IEEE Trans. Inform. Theory}, 56(9):4735--4740, 2010.

\bibitem{kkks}
A.~Ketkar, A.~Klappenecker, S.~Kumar, and P.~K. Sarvepalli.
\newblock Nonbinary stabilizer codes over finite fields.
\newblock {\em IEEE Trans. Inform. Theory}, 52(11):4892--4914, 2006.

\bibitem{KPSW}
A.~Kosovac, J.~Plechinger, N.~Spirovski, and V.~Weger.
\newblock New results on the subfield metric.
\newblock {\em ArXiv 2611.18228}, 2026.

\bibitem{LaG12}
G.~G. La~Guardia.
\newblock Asymmetric quantum {R}eed-{S}olomon and generalized {R}eed-{S}olomon codes.
\newblock {\em Quantum Inf. Process.}, 11(2):591--604, 2012.

\bibitem{Lag2}
G.~G. La~Guardia.
\newblock On the construction of nonbinary quantum {BCH} codes.
\newblock {\em IEEE Trans. Inform. Theory}, 60(3):1528--1535, 2014.

\bibitem{Liu}
H.~Liu, X.~Liu, and Y.~Yuan.
\newblock New asymmetric quantum codes from matrix-product codes.
\newblock {\em Quantum Inf. Process.}, 23(6):Paper No. 229, 2024.

\bibitem{macwilliams}
F.~J. MacWilliams and N.~J.~A. Sloane.
\newblock {\em The theory of error-correcting codes}.
\newblock North-Holland Mathematical Library, Vol. 16. North-Holland Publishing Co., Amsterdam-New York-Oxford, 1977.

\bibitem{Puri}
S.~Puri, L.~St-Jean, J.~A. Gross, A.~Grimm, N.~E. Frattini, P.~S. Iyer, A.~Krishna, S.~Touzard, L.~Jiang, A.~Blais, S.~T. Flammia, and S.~M. Girvin.
\newblock Bias-preserving gates with stabilized cat qubits.
\newblock {\em Sci. Adv.}, 6(34):Paper No. eaay5901, 2020.

\bibitem{Putterman}
H.~Putterman et~al.
\newblock Hardware-efficient quantum error correction via concatenated bosonic qubits.
\newblock {\em Nature}, 638(8052):927--934, 2025.

\bibitem{LDPCcat}
D.~Ruiz, J.~Guillaud, A.~Leverrier, M.~Mirrahimi, and C.~Vuillot.
\newblock {LDPC}-cat codes for low-overhead quantum computing in 2{D}.
\newblock {\em Nat. Commun.}, 16(1):Paper No. 1040, 2025.

\bibitem{sanjose_good_triorthogonal}
R.~San-José.
\newblock Asymptotically good binary triorthogonal codes and higher-level transversal gates.
\newblock {\em ArXiv 2609.08203}, 2026.

\bibitem{asymmetricConstructions}
P.~K. Sarvepalli, A.~Klappenecker, and M.~R\"{o}tteler.
\newblock Asymmetric quantum codes: constructions, bounds and performance.
\newblock {\em Proc. R. Soc. Lond. Ser. A Math. Phys. Eng. Sci.}, 465(2105):1645--1672, 2009.

\bibitem{shorFirstQECC}
P.~W. Shor.
\newblock Scheme for reducing decoherence in quantum computer memory.
\newblock {\em Phys. Rev. A}, 52(4):R2493--R2496, 1995.

\bibitem{cssoriginal2}
A.~M. Steane.
\newblock Multiple-particle interference and quantum error correction.
\newblock {\em Proc. R. Soc. Lond. Ser. A}, 452(1954):2551--2577, 1996.

\bibitem{css2}
A.~M. Steane.
\newblock Simple quantum error-correcting codes.
\newblock {\em Phys. Rev. A}, 54(6):4741--4751, 1996.

\bibitem{TBF}
D.~K. Tuckett, S.~D. Bartlett, and S.~T. Flammia.
\newblock Ultrahigh error threshold for surface codes with biased noise.
\newblock {\em Phys. Rev. Lett.}, 120(5):Paper No. 050505, 2018.

\bibitem{WFLX}
L.~Wang, K.~Feng, S.~Ling, and C.~Xing.
\newblock Asymmetric quantum codes: characterization and constructions.
\newblock {\em IEEE Trans. Inform. Theory}, 56(6):2938--2945, 2010.

\bibitem{Woot}
W.~K. Wootters and W.~H. Zurek.
\newblock A single quantum cannot be cloned.
\newblock {\em Nature}, 299(5886):802--803, 1982.

\bibitem{Zhong}
H.-S. Zhong et~al.
\newblock Quantum computational advantage using photons.
\newblock {\em Science}, 370(6523):1460--1463, 2020.

\end{thebibliography}

\end{document}